\documentclass{article}
\usepackage{amsmath,graphicx,cite,times,fullpage,ifthen}
\usepackage[bf,sf]{caption}
\usepackage[nofiglist,nomarkers]{endfloat}

\def \nmth{n^{\rm th}_m}
\def \ncth{n^{\rm th}_c}
\def \nsigth{n^{\rm th}_\sigma}
\def \nRth{n^{\rm th}_R}

\def \neff{n^{\rm th}_{\rm eff}}

\def \nm{\bar{n}_m}

\def \nmr{n_m^+}
\def \nmb{n_m^-}

\def \SbII{\bar{S}_{II, \rm tot}}
\def \SNII{S^{\rm N}_{II, \rm tot}}

\def \xzp{x_{\rm zp}}

\def \hH{ \hat{\mathcal{H}}}
\def \bse{\begin{subequations}}
\def \ese{\end{subequations}}
\newcommand{\PD}{\phantom{\dag}}

\def \gop{\gamma_{\rm opt}}
\def \gM{\gamma_M}
\def \gtot{\gamma_{\rm tot}}

\def \dP{\delta}
\def \dC{\delta_c}

\title{Observation and interpretation of motional sideband asymmetry in a quantum electro-mechanical device}

\author{A. J. Weinstein$^{1,2}$, C. U. Lei$^{1,2}$, E. E. Wollman$^{1,2}$, J. Suh$^{1,2}$, A. Metelmann$^3$, A. A. Clerk$^3$\& K. C. Schwab$^{1,2}$}

\begin{document}

\makeatletter

\renewcommand\@biblabel[1]{#1.}

\def\@cite#1#2{$^{\mbox{\scriptsize #1\if@tempswa , #2\fi}}$}

\setlength{\parindent}{0.25in}
\setlength{\parskip}{0pt}
\newcommand{\spacing}[1]{\renewcommand{\baselinestretch}{#1}\large\normalsize}
\spacing{1.25}

\def\@maketitle{
  \newpage\spacing{1}\setlength{\parskip}{10pt}
    {\Large\bfseries\noindent\sloppy \textsf{\@title} \par}
    {\noindent\sloppy \@author}
}

\newenvironment{affiliations}{
    \setcounter{enumi}{1}
    \setlength{\parindent}{0in}
    \slshape\sloppy
    \begin{list}{\upshape$^{\arabic{enumi}}$}{
        \usecounter{enumi}
        \setlength{\leftmargin}{0in}
        \setlength{\topsep}{0in}
        \setlength{\labelsep}{0in}
        \setlength{\labelwidth}{0in}
        \setlength{\listparindent}{0in}
        \setlength{\itemsep}{0ex}
        \setlength{\parsep}{0in}
        }
    }{\end{list}\par\vspace{10pt}}

\renewenvironment{abstract}{
    \setlength{\parindent}{0in}
    \setlength{\parskip}{0in}
    \bfseries
    }{\par\vspace{10pt}}

\renewcommand\refname{\vspace{-48pt}\setlength{\parskip}{10pt}}

\newenvironment{addendum}{
    \setlength{\parindent}{0in}
    \begin{list}{Acknowledgements}{
        \setlength{\leftmargin}{0in}
        \setlength{\listparindent}{0in}
        \setlength{\labelsep}{0em}
        \setlength{\labelwidth}{0in}
        \setlength{\itemsep}{0pt}
        \let\makelabel\addendumlabel}
    }
    {\end{list}\normalsize}

\newcommand*{\addendumlabel}[1]{\textbf{#1}\hspace{1em}}

\makeatother

\maketitle

\begin{affiliations}
	\item Applied Physics, Caltech, Pasadena, CA,  91125 USA
	\item Kavli Nanoscience Institute, Caltech, Pasadena, CA 91125 USA
	\item Department of Physics, McGill University, Montreal, QC, H3A 2T8 CA
\end{affiliations}


\begin{abstract}
Quantum electro-mechanical systems offer a unique opportunity to probe quantum noise properties in macroscopic devices, properties which ultimately stem from the Heisenberg Uncertainty Principle. A simple example of this is expected to occur in a microwave parametric transducer, where mechanical motion generates motional sidebands corresponding to the up and down frequency-conversion of microwave photons.  Due to quantum vacuum noise, the rates of these processes are expected to be unequal. We measure this fundamental imbalance in a microwave transducer coupled to a radio-frequency mechanical mode, cooled near the ground state of motion. We also discuss the subtle origin of this imbalance: depending on the measurement scheme, the imbalance is most naturally attributed to the quantum fluctuations of either the mechanical mode or of the electromagnetic field.
\end{abstract}

A fascinating aspect of quantum measurement is that the outcome of experiments and the apparent nature of the object under study depend critically on the properites of both the system and the measurement scheme\cite{Braginsky:quantum}.  An excellent illustration is found when considering measurements of the quantum harmonic oscillator. If measured with an ideal energy detector, the observed signals will demonstrate energy level quantization\cite{Santamore:2004,Clerk:2010PRL}; measured instead with an ideal position detector, no evidence of quantized energy levels are found and the measured signals appear to be that of a very cold, classical oscillator\cite{Roucheleau:2010,Teufel:2011}.  The details of the measurement are as essential to the apparent nature of the system under study, as the properties of the system itself - succinctly expressed by Roy Glauber: ``A photon is what a photodetector detects.'' \cite{Roychoudhuri:2008}

To describe the measured noise of quantum systems, it is often useful to make use of so-called quantum noise spectral densities, which in general are not symmetric functions of frequency: $S_{xx}(-\omega)\neq S_{xx}(+\omega)$, where $S_{xx}(\omega)$ is the spectral density of the observable $x(t)$, defined as the Fourier transform of $\langle \hat{x}(t) \hat{x}(0) \rangle$\cite{Clerk:2010}.  For a quantum harmonic oscillator, the negative and positive frequency sides of this spectral density describes the ability of the system to emit or absorb energy.  In the ground state, there is no ability for the harmonic oscillator to emit energy so that $S_{xx}(-\omega_m)=0$.  It can, however, absorb energy and as a result, $S_{xx}(+\omega_m)=\frac{4}{\gamma_m}\xzp^2$, where $\xzp =\sqrt{\hbar/2m\omega_m}$ is the amplitude of zero point fluctuations for a mechanical oscillator with mass $m$, resonance frequency $\omega_m$, and damping rate $\gamma_m$.  More generally, for a mechanical oscillator in a thermal state with occupation factor, $\nm$, the spectral densities follow $S_{xx}(-\omega_m)=\frac{4}{\gamma_m}\xzp^2\nm$ and $S_{xx}(+\omega_m)=\frac{4}{\gamma_m}\xzp^2(\nm+1)$.\cite{Clerk:2010PRL} This asymmetric-in-frequency motional noise spectrum was first measured in atomic systems prepared in quantum ground states of motion \cite{Diedrich:1989, Jessen:1992, Monroe:1995}, where the motional sideband absorption and fluorescence spectra was detected via photodetection.

Analogous quantum noise effects can also be studied in macroscopic mechanical systems, using electro-mechanical and opto-mechanical devices prepared and probed at quantum limits 
\cite{Teufel:2011,Chan:2011,Purdy:2013,Suh:2013}.
These systems exhibit the Raman-like processes of up and down conversion of photons, resulting from the parametric coupling between mechanical motion and electromagnetic modes of a resonant cavity; the rates of these processes should naturally mirror the asymmetry in the mechanical quantum noise spectral density $S_{xx}(\pm \omega_m)$.
Recent experiments in optomechanics have demonstrated this expected imbalance between up and down converted sidebands\cite{Safavi:2012,Brahms:2012}.  Here, we demonstrate the analogous physics in a quantum circuit, where it is now microwave photons (not optical photons) which probe the mechanical motion.

We also address a subtlety about these measurements which originates from their use of linear detection of the scattered electromagnetic field:  they measure the field amplitude (e.g.~via heterodyne detection).  This is contrast to measurements based employing direct photodetection, where one filters the output light and counts photons associated with a motional sideband.  
Although the predicted and measured motional sideband asymmetry obtained using either detection method are identical \cite{Khalili:2012,Safavi:2012,Brahms:2012}, the interpretation is 
more nuanced when one employs linear field detection.  As discussed by Khalili et al.\cite{Khalili:2012}, the asymmetry in this case can be fully attributed to the detector, namely the presence of a precisely tuned correlation between the backaction noise generated by the measurement device and its imprecision noise (see SI). We provide a simple exposition of this physics using standard input-output theory, which lets us easily track the scattering of incident vacuum fluctuations. 
In the case of linear detection of the cavity output field, the imbalance is naturally attributed to the input electromagnetic field fluctuations (classical and quantum);  the intrinsic 
quantum fluctuations of the mechanical mode contribute equally to the up and down-converted spectrum.  In contrast, in experiments which employ direct photodetection\cite{Diedrich:1989}, the imbalance in the output spectrum (in the absence of thermal electromagnetic noise) is naturally attributed to asymmetric quantum noise of the mechanical motion.  After a brief discussion of these theoretical issues, we present measurements of the imbalance in a microwave-frequency electro-mechanical device.

\paragraph*{Theory}
We begin with the Hamiltonian of our electro-mechanical system,
\begin{equation}\label{Eq.StandHamil}
\hH       =        \hbar \omega_{c} \hat{a}^{\dagger} \hat{a}
		+  \hbar \omega_{m} \hat{b}^{\dagger} \hat{b}
		+  \hbar g_0 \hat{a}^{\dagger}\hat{a} \left(\hat{b}+\hat{b}^{\dagger}\right)
\end{equation}
where $\hat{a}\left(\hat{a}^{\dagger}\right)$ is the annihilation (creation) operator of the microwave resonator mode with frequency $\omega_{c}$, $\hat{b}\left(\hat{b}^{\dagger}\right)$ is the annihilation (creation) operator of the mechanical resonator with frequency $\omega_{m}$, and $g_0$ is the parametric coupling strength between the two modes.

We consider the standard regime of a cavity strongly driven at frequency $\omega_p$, where dissipation is treated as per standard input-output theory\cite{Walls:1995quantum}; we also consider
a two-sided cavity, which corresponds to our experimental setup. Writing the cavity and mechanical fields in terms of their clasical and quantum parts, $\hat{a}=e^{-i \omega_p t} (\bar{a}+\hat{d} )$ and $\hat{b}=\bar{b}+\hat{c}$, we linearize to obtain the following Heisenberg-Langevin equations
$$
\dot{\hat{d}} =-\left(i\Delta+\frac{\kappa}{2}\right)\hat{d}-iG\left(\hat{c}+\hat{c}^{\dagger}\right) 
	- \sum_{\sigma=R,L} \sqrt{\kappa_{\sigma} }\hat{d}_{\sigma, \rm in},
$$
$$
\dot{\hat{c}} =-\left(i\omega_{m}+\frac{\gamma_{m}}{2}\right)\hat{c}-iG(\hat{d}+\hat{d}^{\dagger})-\sqrt{\gamma_{m}}\hat{c}_{\rm in},
$$
where $\Delta=\omega_{c}-\omega_{p}$, $G= g_0 |\bar{a}|$, and $\kappa = \kappa_L + \kappa_R $ ($\gamma_m$) is the microwave (mechanical) resonator damping rate.
The operators $\hat{d}_{\sigma, \rm in}\left(t\right)$, $\hat{c}_{\rm in}\left(t\right)$ describe noise incident on the microwave and mechanical resonator, respectively, and satisfy:
$$
	\left \langle \hat{d}_{\sigma, \rm in}^{\PD} \left(t\right) \hat{d}_{\sigma', \rm in}^{\dagger}\left(t^{\prime}\right)\right \rangle
		=  \left( \nsigth + \alpha \right)   \delta_{\sigma, \sigma'} \delta  \left(t-t^{\prime}\right), \,\,\,\,\,\,
	\left \langle \hat{d}^\dagger_{\sigma, \rm in} \left(t\right) \hat{d}_{\sigma', \rm in}^{\PD}\left(t^{\prime}\right)\right \rangle
		=    \nsigth   \delta_{\sigma, \sigma'} \delta  \left(t-t^{\prime}\right), \,\,\,\,\,\,
$$
$$
	\left \langle \hat{c}_{\rm in}^{\PD} \left(t\right) \hat{c}_{\rm in}^{\dagger}\left(t^{\prime}\right)\right \rangle
		=  \left( \nmth + \beta \right)  \delta  \left(t-t^{\prime}\right), \,\,\,\,\,\,
	\left \langle \hat{c}^\dagger_{\rm in} \left(t\right) \hat{c}_{\rm in}^{\PD}\left(t^{\prime}\right)\right \rangle
		=    \nmth   \delta  \left(t-t^{\prime}\right), \,\,\,\,\,\,
$$
Here, $\nmth$ ($\nsigth$) denotes the amount of thermal fluctuations incident on the mechanical resonator (microwave resonator from port $\sigma$), and $\alpha$, $\beta$ describe the quantum vacuum fluctuations driving the microwave and mechanical resonators, respectively; we have $\alpha = \beta = 1$, consistent with the uncertainty principle and the canonical commutation relation of the noise operators. In what follows, we keep $\alpha$ and $\beta$ unspecified in order to clearly track the contributions of both mechanical and electromagnetic vacuum noise to the measured noise spectrum.

We further specialize to the case where a single microwave cavity drive is applied at $\omega_p=\omega_c - \Delta$ with $\Delta$ either $\pm \omega_m$, and consider the up- and down- converted sidebands generated by the mechanical motion.  For simplicity, we ignore any internal loss of the cavity, consider the system to be in the sideband resolved regime ($\kappa \ll \omega_m$), and also consider the limit of a weak cooperativity $4G^2/\kappa \ll \gamma_m$.  This last condition implies that the backaction effects on the mechanics are minimal: the mechanical linewidth and temperature are set by its coupling to its intrinsic dissipative bath.

For amplitude detection either with a linear amplifier as in this experiment, or optical heterodyne detection\cite{Safavi:2012,Brahms:2012}, the symmetric noise spectrum is:
\begin{equation}
\SbII [\omega]= \frac{1}{2} \int dt \left\langle \hat{I}_{\rm tot}\left(t\right) \hat{I}_{\rm tot}\left(0\right) + \hat{I}_{\rm tot}\left(0\right) \hat{I}_{\rm tot}\left(t\right) \right\rangle e^{i\omega t}.
\label{symDet}
\end{equation}
with the amplitude of the output field $\hat{I}_{\rm tot}=\hat{d}_{R,\rm out}^{\PD}+\hat{d}^\dagger_{R,\rm out}$ and where $\hat{d}_{R,\rm out}=\hat{d}_{R,\rm in}+\sqrt{\kappa_R} \hat{d}$. The output spectrum near the cavity resonance for the two choices of drive detuning are found to be
\begin{equation}\label{Eq.SymSpectraRed}
	\SbII [\omega]\Big|_{\Delta=+\omega_{m}} =
		\bar{S}_0 +
		\frac{\kappa_R}{\kappa}\frac{\gop \gamma_{m}}{\left(\frac{\gamma_m}{2}\right)^{2}+
	\left(\omega-\omega_{c}\right)^{2}}
	\left[  \left(\nmth +\frac{\beta}{2}\right) - \left(\neff + \frac{\alpha}{2}\right) \right],
\end{equation}
\begin{equation}\label{Eq.SymSpectraBlue}
	\SbII [\omega]\Big|_{\Delta=-\omega_{m}} =
	\bar{S}_0 +
		\frac{\kappa_R}{\kappa}\frac{\gop \gamma_{m}}{\left(\frac{\gamma_m}{2}\right)^{2}+\left(\omega-\omega_{c}\right)^{2}}
		\left[  \left(\nmth +\frac{\beta}{2}\right) + \left(\neff + \frac{\alpha}{2}\right) \right].
\end{equation}

where for $\Delta = \omega_m$ ($\Delta = -\omega_m$), the up- (down-) converted sideband is centered on the cavity resonance.  The noise floor for both cases is given by
$\bar{S}_0 = \alpha/2 +  n^{\rm th}_R + 4  \kappa_R (n^{\rm th}_c - n^{\rm th}_R) / \kappa$, and we have defined $\neff = 2 \ncth - n^{\rm th}_R$ (where
$\ncth = (\kappa_L n^{\rm th}_L + \kappa_R n^{\rm th}_R) / \kappa$ is the effective cavity thermal occupancy).  
In Fig.1(c), we illustrate the underlying components of this spectrum.

One sees explicitly that the sideband imbalance,  $\SbII [\omega]|_{\Delta=-\omega_{m}}-\SbII[\omega]|_{\Delta=+\omega_{m}}$, is proportional to 
$(2 \neff + \alpha)$, and hence is entirely due to fluctuations in the microwave fields driving the cavity.  This is true both when this noise is thermal, and when it is purely quantum 
(i.e.~$n^{\rm th}_R = n^{\rm th}_L = 0$).  These terms in the spectrum result from the interference between the two ways the incident field noise can reach the output:  either by directly being transmitted through the cavity, or by first driving the mechanical resonator whose position then modulates the amplitude quadrature of the outgoing microwaves (see SI for further insights
based on a scattering approach).
This is the basic mechanism of noise squashing, which in the case of thermal noise was previously observed in a cavity electromechanical system \cite{Roucheleau:2010}.  This mechanism
can also be fully described using a general linear measurement formalism\cite{Khalili:2012}, where it is attributed to the presence of correlations between the backaction and imprecision noises of the detector, correlations which are out-of-phase and have magnitude $\hbar/2$ in the zero-temperature limit.  Interestingly, this precise value plays a special role in the theory
of quantum limits on linear amplification \cite{Clerk:2010} (see SI for more details).  

The above calculation also shows that both thermal and zero-point force noise emanating from the mechanical bath (i.e.~terms $\propto \nmth + \beta/2$) contribute symmetrically to Eqs.~(\ref{Eq.SymSpectraRed}) and (\ref{Eq.SymSpectraBlue}), and hence play no role in determining the asymmetry of the sidebands.  In the weak-cooperativity limit, it is the mechanical bath which almost entirely determines the mechanical oscillator fluctuations.  This suggests that the sideband asymmetry observed using 
linear detection of the scattered field is not directly probing the asymmetric quantum noise spectrum of the mechanical mode.  

In contrast, direct measurement of the sideband signal via photon counting yields the normal ordered spectrum,
\begin{equation}
\SNII\left(\omega\right)=\int dt\left\langle: \hat{I}_{\rm tot}\left(t\right)\hat{I}_{\rm tot}\left(0\right):\right\rangle e^{i\omega t},
\label{photoDet}
\end{equation}
with output spectra given by
\begin{align}
	\SNII[\omega]\Big|_{\Delta=+\omega_{m}} =&
		\left( \bar{S}_0 - \frac{\alpha}{2} \right) +
			\frac{\kappa_R}{\kappa} \frac{\gop \gamma_{m}}{\left(\frac{\gamma_m}{2}\right)^{2}+\left(\omega-\omega_{c}\right)^{2}}\left( \nmth-\neff \right),
\\
	\SNII [\omega]\Big|_{\Delta=-\omega_{m}} =&
		\left( \bar{S}_0 - \frac{\alpha}{2} \right) +
			\frac{\kappa_R}{\kappa} \frac{\gop \gamma_{m}}{\left(\frac{\gamma_m}{2}\right)^{2}+\left(\omega-\omega_{c}\right)^{2}}\left( \nmth+\beta + \neff \right).
\end{align}
Note that when one sets $\alpha = \beta = 1$, the asymmetry of these normal-ordered spectra, $\SNII[\omega]|_{\Delta=-\omega_{m}}-\SNII[\omega]|_{\Delta=+\omega_{m}}$, is {\it identical} to that
obtained from the linear measurement (where spectra are calculated using Eq.~(\ref{symDet})).  
In this case, however, the asymmetry is naturally attributed to both the mechanical quantum fluctuations, $\beta$, and to the thermal microwave fluctuations
described by $\neff$; this is illustrated in Fig.1(b).  Note that in direct photodetection, one cannot attribute the zero-temperature sideband asymmetry to a correlation between backaction-driven position fluctuations and imprecision noise, as there is no imprecision noise floor.

While the above simple calculations suggest that the sideband asymmetry measured using linear detection versus direct photodetection have different origins, it is no accident that the 
magnitudes of the asymmetry are the same in both schemes.  This follows directly from the fact that the canonical commutation relation of the output field is the same as
the input field, $\left[\hat{d}_{R,\rm out}^{\PD}[\omega],\hat{d}^{\dagger}_{R,\rm out}[\omega'] \right]=\alpha \delta(\omega+\omega')$.  It necessarily follows that the spectra in Eqs.~(\ref{symDet}) and Eqs.~(\ref{photoDet}) will differ only by a frequency-independent noise floor of magnitude $\alpha/2$ \cite{Khalili:2012}.  If one assumes this commutation relation, then one can legitimately say that both spectra essentially measure the same thing.  However, on a formal level, this involves an additional assumption on the value of $\beta$:  (if $\beta \neq \alpha$, then the output commutator would not be the same as the input, see SI).

Having explored the interpretation subtleties associated with sideband asymmetry, we now turn to presenting our main result: the experimental observation of this imbalance
in a microwave-cavity based electromechanical system.

\paragraph*{Experiment}

Our system is composed of a superconducting microwave resonator, also referred to as ``cavity'', where the resonance frequency is modulated by the motion of a compliant membrane \cite{Suh:2013}. This frequency modulation leads to the desired parametric coupling between microwave field and mechanical motion (Fig.2(a)). Measurements of the cavity response below 100 mK yield the resonance frequency $\omega_c=2\pi \times 5.4$ GHz, total loss rate $\kappa=2\pi \times 860$ kHz, output coupling rate $\kappa_R=2\pi \times 450$ kHz, and input coupling rate $\kappa_L=2\pi \times 150$ kHz.  The capacitor top gate is a flexible aluminum membrane (40$\mu$m$\times$40$\mu$m$\times$150nm) with a fundamental drumhead mode with resonance frequency $\omega_m=2\pi \times 4.0$ MHz and intrinsic loss rate $\gamma_m=2\pi \times 10$ Hz at 20mK.  Motional displacement of the top gate modulates the microwave resonance frequency with an estimated coupling rate of $g_0 = \frac{\partial \omega_c}{\partial x} \xzp = 2\pi \times 16$ Hz.

In Fig. 2(c), we present a schematic of the measurement circuit.  Tunable cavity filters at room temperature reduce the source phase noise to the thermal noise at 300K; cryogenic attenuators further reduce the noise down to the shot noise level\cite{Roucheleau:2010}. A pair of microwave switches at the device stage select between the device or a bypass connection for high precision noise floor calibration of the cryogenic amplifier.  The output signal passes through two cryo-circulators at $\sim$100mK followed by a cryogenic low-noise amplifier at 4.2K, and finally to a room temperature circuits for analysis. The occupation factor of the microwave resonator, $\ncth$, which is expected to thermalize below $5\times10^{-3}$ at temperatures below 50mK, can be increased and controlled by the injection of microwave frequency noise from amplified room temperature Johnson noise. From careful measurements of the noise power emanating from the cavity at zero pumping and comparing this to power spectra with the bypass switched in place  (see SI), we conclude that there is a small contribution to $\ncth$ due to thermal radiation from the isolated port of the cryogenic circulators, given by the occupation factor $\nRth= 0.34\pm0.03$.

When a single microwave tone is applied to the device at $\omega_p$, the parametric coupling converts mechanical oscillations at $\omega_m$ to up and down-converted sidebands at $\omega_p\pm\omega_m$.  In this experiment, we apply microwave tones at frequencies near $\omega_c\pm\omega_m$ and at powers given by the mean number of photons in the resonator, $n_p$.  The microwave resonance suppresses motional sidebands outside of the linewidth and we consider only the contributions of signals converted to frequencies near $\omega_c$.  These are the Lorentzian components of the noise power spectra of Eqs. (\ref{Eq.SymSpectraRed}) and (\ref{Eq.SymSpectraBlue}), which for the remainder of the paper are denoted by ``+'' and ``-'', respectively, and are labeled in Fig.1(c).

Throughout the measurement, we simultaneously apply three microwave tones. We place a cooling tone at $\omega_c-\omega_m-\dC$ to control the effective mechanical damping rate, $\gM$, and mode occupation, $\nm$, via back-action cooling\cite{Marquardt:2007}. Two additional probe tones, placed at $\omega_c\pm(\omega_m+\dP)$, produce up and down converted sidebands symmetrically detuned from cavity center (Fig.3(a)).  The detunings are chosen to ensure no interference between the sidebands ($\dC=2 \pi \times 30$ kHz, $\dP = 2 \pi\times 5$ kHz) so that we may consider the probe sidebands as independent measurements of the dressed mechanical mode as validated by theory.

To summarize the main differences between the simplified theory model presented above and our actual experiment, we measure the mechanical sidebands produced in a two-port microwave resonator with limited sideband resolution and a noisy output port, and in the presence of multiple injected tones with a range of detunings and powers. From further analysis (see SI), we estimate corrections to the sideband asymmetry that are $\ll 1$ and far below the measurement resolution of our system. 

To convert the motional sideband powers into equivalent mechanical occupation, we turn off the cooling tone and measure the probe sidebands ($\dP=2\pi \times 500$ Hz) with low optical damping ($n_p^+ = n_p^- \simeq 5\times10^2$) and high mechanical occupation set by the cryostat temperature. Regulating the temperature to calibrated levels between 20 to 200mK, we calculate the integrated noise power under the sideband Lorentzians, $P_m^\pm$, normalized by the respective microwave probe power transmitted through the device, $P_{\rm thru}^\pm$. In the limit of high thermal occupation, the normalized power is directly proportional to $\nm$.\cite{Hertzberg:2009}  As we vary the cryostat temperature, $T$, we compare the normalized power to the thermal occupation factor $[\exp(\frac{\hbar\omega_m}{k_BT})-1]^{-1}$ (Fig.2(b)).  A linear fit yields the conversion factors for the up-converted ($\nmr$) and down-converted ($\nmb$) sidebands: $\nmr=(9.9\pm0.2) \times 10^8 \cdot P^+_m/P^+_{\rm thru}$ and $\nmb=(5.4\pm0.1) \times 10^8 \cdot P^-_m/P^-_{\rm thru}$.

Further detuning the probe tones ($\dP=2\pi \times 5$ kHz) and turning on the cooling tone ($\dC=2\pi \times 30$ kHz), we explore the sideband ratio, $\nmb/\nmr$, over various the mechanical and microwave occupations.  To reduce $\nm$ to values approaching 1, we increase the cooling tone power up to $n_p^{\rm cool} = 4 \times 10^5$.  For sideband characterization, the probe tone powers are set to $n^{-}_p=n^{+}_p = 10^5$ and the probe sideband spectra are analyzed using the conversion factors described above.  The imbalance between $\nmb$ and $\nmr$ is clearly evident in the noise spectra (Fig.3(b)).

As further demonstration of the asymmetry with respect to $\neff$, we plot $\nmb/\nmr$ as a function of $\nmr$ in Fig. 3(c).  Each curve corresponds to one setting of injected microwave noise. The data shows excellent agreement to the expected ratio, $\nmb/\nmr=1 + (2\neff+1)/\nmr$. This relationship highlights the combined effect of quantum and classical noise in Eqs.~(\ref{Eq.SymSpectraRed}) and ~(\ref{Eq.SymSpectraBlue}) (see SI).  By fitting each curve to a two parameter model: $a+b/\nmr$, we find an average constant offset $a=0.99\pm0.02$ for all curves, accurately matching the model and confirming our calibration techniques. Fitting for $b$, the data indicates $\neff$ spanning $0.71$ to $4.5$ with uncertainty all within $\pm0.09$ quanta.

To quantify the contributions due to quantum fluctuations and classical cavity noise, we fix the cooling tone power at $n_p^{\rm cool} = 4 \times 10^5$ ($\gM=2 \pi \times 360$ Hz) and measure the imbalance $\nmb-\nmr$ as we sweep $\neff$.  At each level, we measure the average noise power density, $\eta$, over a $250$ Hz window centered at $\omega_c$ and away from any motional sideband. Over this range, $\eta$ contains two contributions: the noise radiating out of the microwave resonator, proportional to $\neff$, and the detector noise floor, set by the noise temperature of the cryogenic amplifier ($T_N\approx3.6$K). We directly measure the detector noise floor by switching from the device to an impedance-matched bypass connection and measure the noise power density, $\eta_0$, over the same window with matching detected tone powers.

In Fig. 3(d), we plot the sideband imbalance against the noise floor increase, $\Delta\eta=\eta-\eta_0$, which is expected to follow: $\nmb-\nmr=2\neff+1 = 4\lambda\cdot\Delta\eta+1$, where $\lambda$ is the conversion factor for $\Delta\eta$ in units of cavity quanta, $\ncth$.  The data clearly follows a linear trend with a slope of $\lambda=(2.7\pm0.1)\times10^{-1}$ (aW/Hz)$^{-1}$.  More importantly, we observe an offset of $1.2\pm0.2$, in excellent agreement with the expected quantum imbalance of ``+1'' from the quantum fluctuations of the microwave field.

As an additional check, we also consider the sideband average, $(\nmr+\nmb)/2$, as a function of $\Delta\eta$.  Averaging Eqs.~(\ref{Eq.SymSpectraRed}) and ~(\ref{Eq.SymSpectraBlue}), we see that the resulting occupation, $\nm+\frac{\beta}{2}$, does depend on $\neff$ due to the coupling between the mechanical and microwave modes, $\nm=\frac{\gamma_m}{\gtot}\nmth+\frac{\gop}{\gtot}(2\ncth+ \alpha) + \frac{\gop^{\rm cool}}{\gtot}\ncth$, where $\gop$ ($\gop^{\rm cool}$) is the optical coupling rate for the individual probe (cooling) tones.  Accounting for this so-called back-action heating of the mechanical mode\cite{Marquardt:2007, Suh:2013}, we recover $\lambda=(2.5\pm0.2)\times10^{-1}$ (aW/Hz)$^{-1}$, consistent with the imbalance results above.

Notably, the average sideband occupation does contain contributions from mechanical zero-point fluctuations. Future experiments could infer the mechanical quantum contribution of $\frac{\beta}{2}$ with a method to independently calibrate $\nm$ to high accuracy, for example, with a passively cooled high frequency mechanical mode thermalized to a primary low temperature thermometer. 

In summary, we report the quantum imbalance between the up and down-converted motional sideband powers in a cavity electro-mechanical system measured with a symmetric, linear detector. We show that for linear detection of the microwave field, the imbalance arises from the correlations between the mechanical motion and the quantum fluctuations of the microwave detection field.  For normal-ordered detection of the microwave field, however, the imbalance arises directly from the quantum fluctuations of the mechanics. By further assuming that the output microwave field satisfies the cannonical commutator, which also determines the quantum fluctuations of the mechanical mode, the measurement can be interpreted as performing either symmetric or normal-ordered detection regardless of the type of detector utilized. In both scenarios, the imbalance in motional sidebands is a fundamental quantity originating from the Heisenberg's uncertainty relations and provides a quantum calibrated thermometer for mesoscopic mechanical systems.

\begin{addendum}
 \item[Acknowledgements] We would like to acknowledge Yanbei Chen and Matthew Woolley for helpful discussion. This work is supported by funding provided by the Institute for Quantum Information and Matter, an NSF Physics Frontiers Center with support of the Gordon and Betty Moore Foundation (NSF-IQIM 1125565), by DARPA (DARPA-QUANTUM HR0011-10-1-0066), by NSF (NSF-DMR 1052647, NSF-EEC 0832819), and by the DARPA ORCHID program under a grant from AFOSR.
 \item[Competing Interests] The authors declare no
competing financial interests.
 \item[Correspondence] Correspondence and requests for material
should be addressed to Keith Schwab~(email: schwab@caltech.edu).
\end{addendum}

\begin{figure}
\centering\includegraphics[scale=0.8]{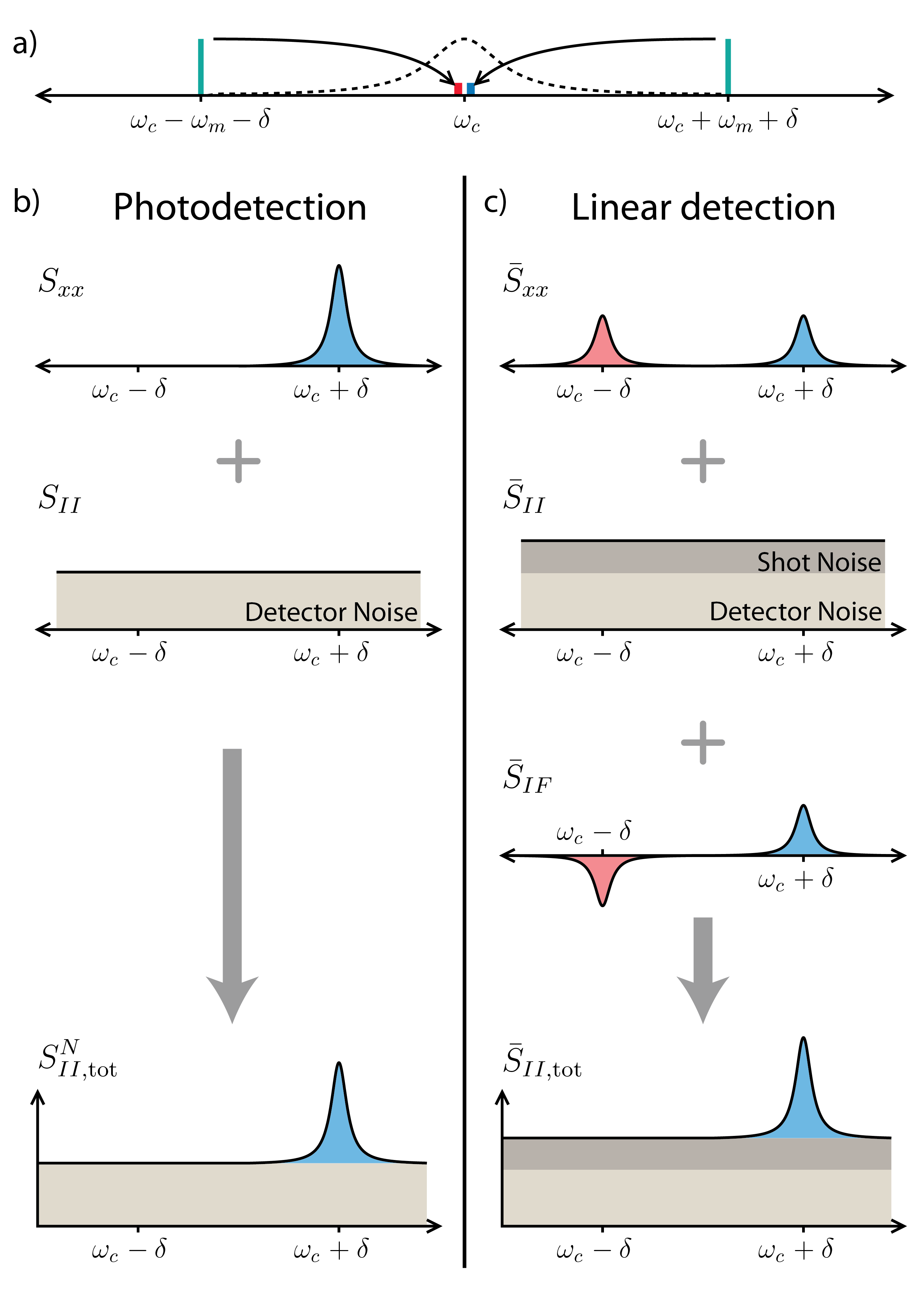}
\caption{Comparison between photodetection and linear detection. \textbf{a.} Pump scheme.  We consider a single microwave cavity (dotted line) pumped at $\omega_c \pm (\omega_m + \dP)$ (green).  The up-converted (red) and down-converted (blue) motional sidebands are placed tightly within the cavity linewidth.  For figure clarity, the occupation of the microwave and mechanical modes are assumed to be zero. \textbf{b.} Normal-ordered detection. Photodetection is sensitive to the asymmetric motional noise spectrum, $S_{xx}$. The photodetector is not sensitive to microwave shot noise and the noise floor ($S_{II}$) is from detector non-idealities (light grey), analogous to dark counts for a photodetector. \textbf{c.} Linear detection.  The contribution from the symmetrized motional noise, $\bar S_{xx}$, is present in both sidebands.  Microwave shot noise (dark grey) and amplifier noise (light grey) combine to form the imprecision noise $\bar S_{II}$.  This measurement is sensitive to noise correlation between the microwave and mechanical modes ($\bar S_{IF}$), which results in asymmetric sqashing (red) and anti-squashing (blue) of the noise floor.  Though the source is different, the sideband imbalance is identical in both photodetection and linear detection.  For mathematical description of $S_{II}$ and $S_{IF}$, refer to SI.}
\end{figure}

\begin{figure}
\centering\includegraphics{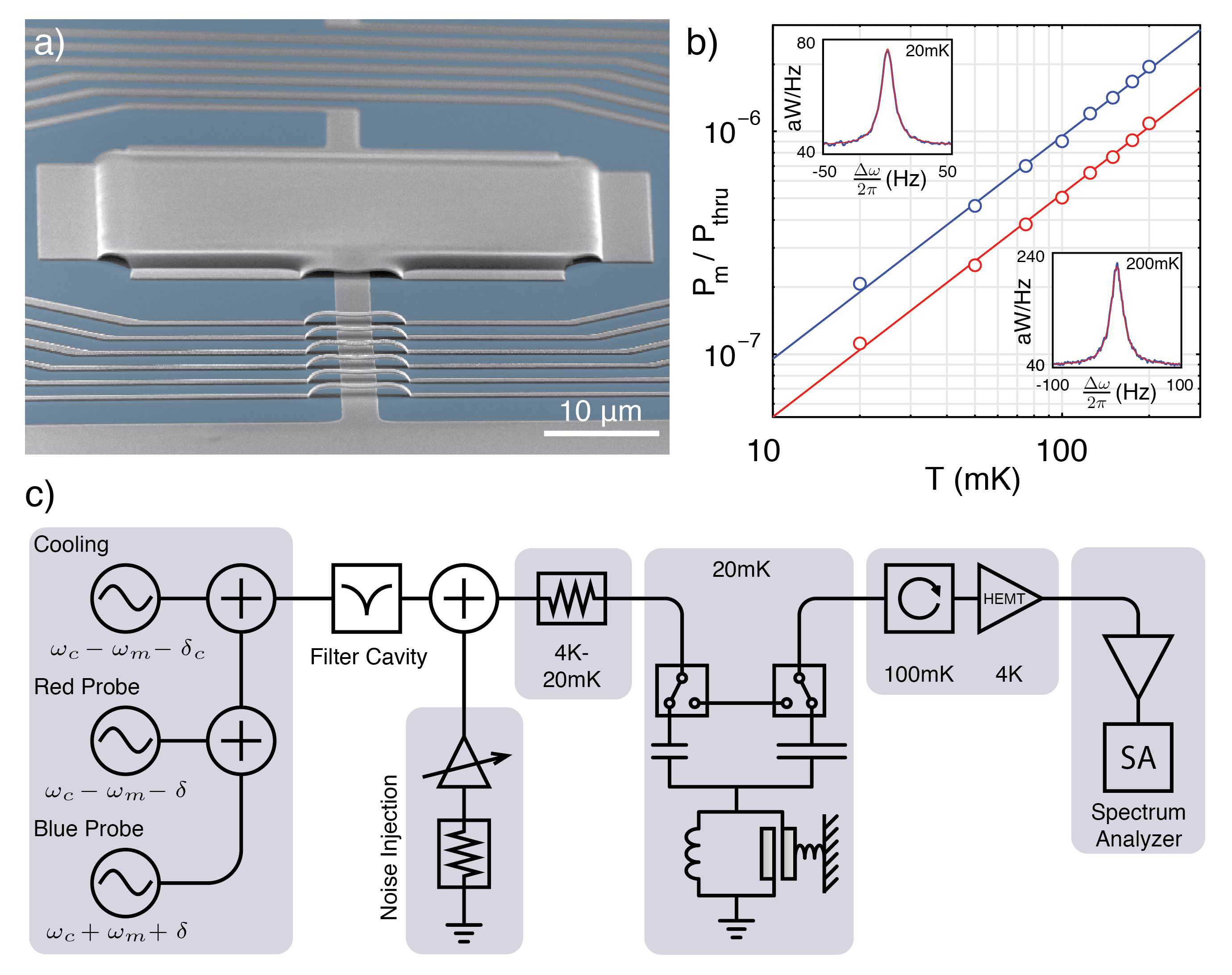}
\caption{Device, calibration, and measurement scheme. \textbf{a.} Electron micrograph of the measured device. A suspended aluminum (grey) membrane patterned on silicon (blue) forms the electro-mechanical capacitor. It is connected to the surrounding spiral inductor to form a microwave resonator.  Out of view, coupling capacitors on either side of the inductor couple the device to input and output co-planar waveguides. \textbf{b.} Motional sideband calibration.  The cryostat temperature is regulated while the  mechanical mode is weakly probed with microwave tones set at $\omega_c+\omega_m+\dP$ (blue) and at $\omega_c-\omega_m-\dP$ (red) detunings, with $\dP=2 \pi \times 500$ Hz.  The observed linear dependence provides the calibration between the normalized sideband power and the mechanical occupation factor. Inset, up-converted motional sideband spectra collected at 20mK (top) and 200mK (bottom), with $\Delta \omega = \omega-(\omega_c-\dP)$. \textbf{c.} Schematic of the microwave measurement circuit.}
\end{figure}

\begin{figure}
\centering\includegraphics[scale=0.9]{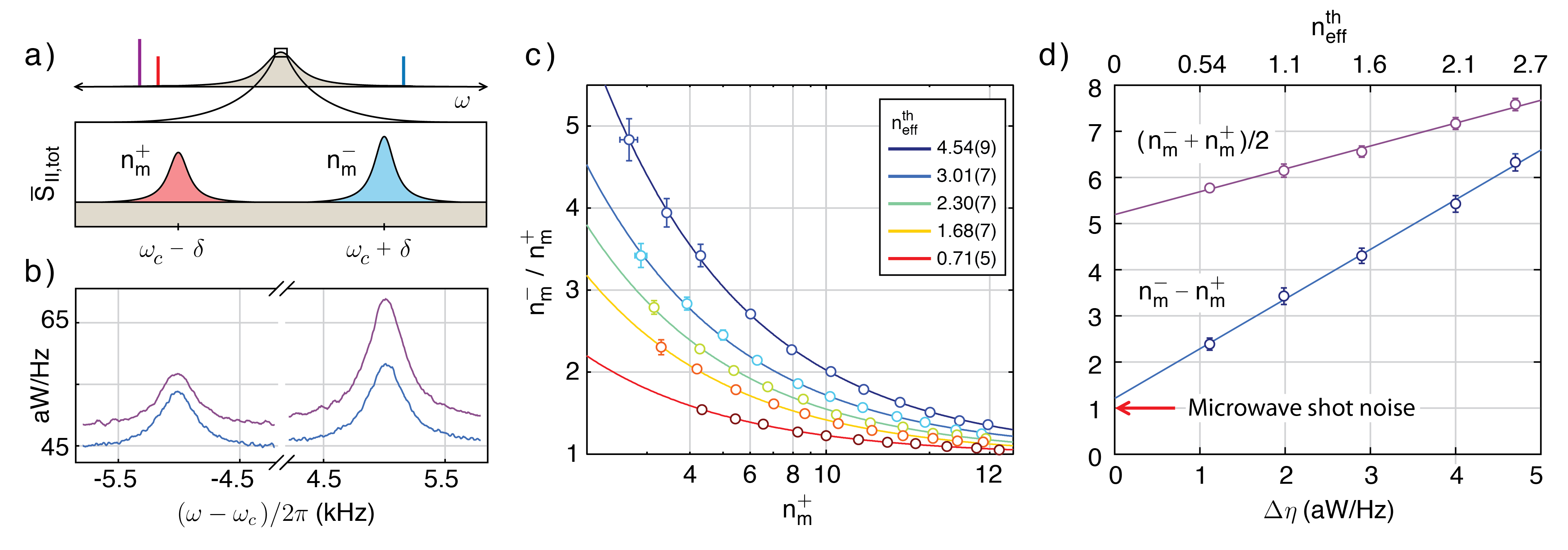}
\caption{Sideband asymmetry. \textbf{a.} Pump scheme.  Three tones are placed about the microwave resonance.  Two probe tones generate up-converted (red) and down-converted (blue) sidebands.  An additional tone (purple) cools the mechanical mode. \textbf{b.}  Sideband spectra.  $\SbII(\omega)$ measured at $\neff=0.60$ (blue) and $2.5$ (purple) with $\nm=4.7\pm0.1$. \textbf{c.}  Sideband asymmetry.  The ratio $\nmb/\nmr$ vs. $\nmr$ is plotted for increasing noise injection. \textbf{d.} Sideband imbalance (blue) and sideband average (purple) vs. the measured noise increase, $\Delta\eta$. Sideband imbalance, $\nmb-\nmr$, and average, $(\nmb+\nmr)/2$, exhibit a linear trend with $\Delta\eta$. The imbalance at $\Delta\eta = 0$ is the quantum imbalance due to the squashing of fluctuations of the microwave field.}
\end{figure}

\end{document}


\maketitle

\section{Input-Output Theory}

In this section, we give a framework to calculate the output noise spectrum of an opto/electro-mechanical system with arbitrary pump configuration by utilizing the input-output theory. As a first example, we analyze an ideal (without intrinsic losses) two-port opto/electro-mechanical system with a single pump tone either at frequency $\omega_{p}=\omega_{c} - \omega_{m}$ or $\omega_{p}=\omega_{c} + \omega_{m}$, and discuss the origin of the sideband asymmetry in the output noise spectrum. We then use this method to study the system in our experiment, i.e., a two-port electro-mechanical system with three pumps (balanced detuned two tones and a cooling tone).

We start with the standard Hamiltonian of an opto/electro-mechanical system 
\begin{equation}\label{Eq.StandHamil}
\hH       =        \hbar \omega_{c} \hat{a}^{\dagger} \hat{a}
		+  \hbar \omega_{m} \hat{b}^{\dagger} \hat{b}
		+  \hbar g_0 \hat{a}^{\dagger}\hat{a} \left(\hat{b}+\hat{b}^{\dagger}\right)
		+  \hH_{\rm drive} + \hH_{\rm diss},
\end{equation}
where $\hat{a}$ ($\hat{a}^{\dagger}$) is the annihilation (creation) operator of the cavity field. $\hat{b}$ ($\hat{b}^{\dagger}$) is the annihilation (creation) operator of the phonon, $g_0$ is the coupling strength between the cavity and the mechanical oscillator. 
We assume an external driving, described by $\hH_{\rm drive}$, which is applied on the input port on the left side of the cavity. The optical and the mechanical system are both coupled to dissipative baths, described by $\hH_{\rm diss} $, giving rise to the decay rates $\gamma_{m}$ for the mechanical and $\kappa$ for the optical system. 
The total cavity linewidth $\kappa$ consists of the contributions from the different decay channels, namely the right ($R$) and the left ($L$) port, as well from intrinsic losses ($I$) inside of the cavity, i.e., $\kappa = \kappa_{R}+\kappa_{L}+\kappa_{ I}$.

For large pumping fields, we may split the fields into classical and quantum components, $\hat{a}\rightarrow \bar{a} + \hat{d}$ and $\hat{b} \rightarrow \bar{b} + \hat{c}$, where $\hat{d}$ and $\hat{c}$ describe the quantum fluctuations of the cavity photon and the phonon. By using input-output theory and neglecting the second order contributions from the quantum fluctuations, the linearized quantum Langevin equations are
\bse
\begin{align} 
\dot{\hat{d}} & = - i \omega_{c}^{\prime}\hat{d} - \frac{\kappa}{2}\hat{d} - ig_0 \ \bar{a}(t)   \left[\hat{c}+\hat{c}^{\dagger}\right]
		  - \sum_{\sigma \in {L,R,I}}\sqrt{\kappa_\sigma} \hat{d}_{\sigma ,\rm in},
\\
\dot{\hat{c}} &=  - i \omega_{m}\hat{c}  - \frac{\gamma_{m}}{2} \hat{c} - ig_0 \left[\bar{a}^{\ast}(t)\hat{d}+\bar{a}\left(t\right)\hat{d}^{\dagger}\right]
		  -\sqrt{\gamma_{m}}\hat{c}_{\rm in},
\end{align}\label{eq:EQM}
\ese
where $\omega_{c}^{\prime}=\omega_{c}+g(\bar{b}+\bar{b}^{*})\simeq\omega_{c}$. Including the possibility of multiple drives at frequencies $\omega_n$, we obtain
$\bar{a}(t)=\sum_{n}\bar{a}_{n}e^{-i\omega_{n}t}$ as the driving field inside the cavity, with $\bar{a}_n=\frac{\sqrt{\kappa_L}\alpha_n}{\frac{\kappa}{2}-i(\omega_n -\omega_c)}$. Without loss of generality, we take $\bar{a}_n$ to be real.
In Equations~(\ref{eq:EQM}), $\hat{d}_{\sigma,\rm in}$ describes the input fluctuations to the cavity from channel $\sigma$ with damping rate $\kappa_{\sigma}$, and $\hat{c}_{\rm in}$ describes the input fluctuations to the mechanical oscillator. 
The input field operators satisfy the following commutation relations
\begin{equation}\label{Eq.correlators}
\begin{array}{cc}
\left[ \hat{d}_{\sigma,\rm in}^{\PD} (t), \hat{d}_{\sigma^{\prime}\rm in}^{\dagger} (t^{\prime})\right] 
			= \alpha_{\sigma} \delta_{\sigma \sigma^{\prime}} \ \delta \left(t-t^{\prime}\right),
\ \  & 
\left[\hat{c}_{\rm in}^{\PD}(t),\hat{c}_{\rm in}^{\dagger}(t^{\prime})\right]          
			= \beta  \ \delta\left(t-t^{\prime}\right),
\\[0.2cm]
\left\langle \hat{d}_{\sigma,\rm in}^{\dagger}(t)\hat{d}_{\sigma^{\prime}\rm in}^{\PD}(t^{\prime})\right\rangle 
			= n_{\sigma}^{\rm th} \delta_{\sigma \sigma^{\prime}} \ \delta\left(t-t^{\prime}\right),
\ \  & 
\left\langle \hat{c}_{\rm in}^{\dagger}(t)\hat{c}_{\rm in}^{\PD}(t^{\prime})\right\rangle     =  
			\nmth   \  \delta\left(t-t^{\prime}\right),
\end{array}
\end{equation}
where $\alpha_{\sigma}=\beta=1$, $n_{\sigma}^{\rm th}$ is the photon occupation in port $\sigma$, and $\nmth=1/(\exp(\frac{\hbar\omega_{m}}{k_{B}T})-1)$ is the thermal occupation factor of the bath responsible for the intrinsic mechanical dissipation. The total thermal occupation of the cavity is the weighted sum of the contributions from different channels, $\ncth =\sum_{\sigma}\frac{\kappa_{\sigma}}{\kappa}n_{\sigma}^{\rm th}.$
Note, that the relations in Eq.~(\ref{Eq.correlators}) are only valid if we deal with frequencies close to cavity resonance. 

\subsection*{Single Tone}

We start with the case of a single pump tone at frequency $\omega_p = \omega_c - \Delta$, where the drive detuning $\Delta$ is chosen to either be $\pm \omega_m$; 
our goal is to make the origin of the asymmetry between the spectra measured for these two cases clear.  For maximum clarity,  we also consider the good-cavity limit  $\omega_m \gg \kappa$ and work within the rotating-wave approximation.  In this limit, we can describe the relevant spectra in terms of a $3 \times 3$ scattering matrix, involving the fields $ \mathbf{\hat D^{+(-)}} \equiv (\hat d_{R} ,\hat d_{L}, \hat c^{(\dag)}  )^{\rm T}$, where the $+(-)$ refers to a driving on the red (blue) sideband, i.e., $\Delta = \pm \omega_m$. 
By using the input-output relations $\hat d_{\sigma,\rm out} = \hat d_{\sigma,\rm in} + \sqrt{\kappa_\sigma} \hat d_\sigma $ and $\hat c_{\rm out} = \hat c_{\rm in} + \sqrt{\gamma_m} \hat c $ and solving the corresponding quantum Langevin equations, we obtain in frequency space (working in a rotating frame at the cavity frequency)
\footnote{
\[
\begin{array}{cc}
\hat{f}\left(\omega\right)=\int_{-\infty}^{\infty}dt\,\hat{f}(t)e^{i\omega t},\ \  & \hat{f}^{\dagger}\left(\omega\right)=\int_{-\infty}^{\infty}dt\,\hat{f}^{\dagger}(t)e^{i\omega t}=\left[\hat{f}\left(-\omega\right)\right]^{\dagger}.\end{array}
\]
}
\begin{align}
\mathbf{\hat D}_{\rm out}^{\pm}[\omega] &= \mathbf{s}^{\pm}[\omega] \mathbf{\hat D}_{\rm in}^{\pm}[\omega].
\end{align}
For frequencies close to the cavity resonance (i.e.,~$|\omega - \Delta| \ll \kappa$), the scattering matrix $\mathbf{s}[\omega]$ is
\begin{equation}\label{Eq.ScatteringMatrix}
 \mathbf{s}^{\pm}[\omega] =
\left(
\begin{array}{ccc}
	  \displaystyle  \left[1 - \frac{2\kappa_{ R}}{\kappa} \pm \frac{\kappa_{ R}}{\kappa}
				   \frac{  \gop}
					{N^{\pm}[\omega]} \right]
	& \displaystyle    \left[- \frac{2\sqrt{\kappa_{L} \kappa_{ R}} }{\kappa}
				 \pm \frac{\sqrt{\kappa_{L} \kappa_{ R}} }{\kappa}
				   \frac{  \gop}
				        {N^{\pm}[\omega]} \right]
	& \displaystyle            \sqrt{ \frac{\kappa_{ R}}{\kappa} }
				   \frac{i \sqrt{\gamma_m \gop} }
				        {N^{\pm}[\omega] } 
\\[0.5cm]
	  \displaystyle          \left[- \frac{2\sqrt{\kappa_{ L} \kappa_{ R}} }{\kappa}
				 \pm \frac{\sqrt{\kappa_{ L} \kappa_{ R}} }{\kappa}
				   \frac{  \gop}
				        {N^{\pm}[\omega]} \right]
	& \displaystyle  \left[1 - \frac{2\kappa_{ L}}{\kappa} \pm \frac{\kappa_{ L}}{\kappa}
				   \frac{  \gop}
					{N^{\pm}[\omega]} \right]
	& \displaystyle            \sqrt{ \frac{\kappa_{L}}{\kappa} }
				   \frac{i \sqrt{\gamma_m \gop} }
				        {N^{\pm}[\omega]}   
\\[0.5cm]
	  \displaystyle            \sqrt{ \frac{\kappa_{ R}}{\kappa} }
				   \frac{i \sqrt{\gamma_m \gop} }
				        {N^{\pm}[\omega]}  
	& \displaystyle            \sqrt{ \frac{\kappa_{ L}}{\kappa} }
				   \frac{i \sqrt{\gamma_m \gop} }
				        {N^{\pm}[\omega]}  
	& \displaystyle  \left[1 - \frac{\gamma_m }
				        {N^{\pm}[\omega]} \right] \\
\end{array}
\right).
\end{equation}
Here, the denominator $N^{\pm}[\omega]$ describes the mechanical response including optical damping / anti-damping:
\begin{align}\label{Eq.denominator}
N^{\pm}[\omega] =& -i(\omega \mp \omega_m) + \frac{\gamma_m \pm \gop}{2},
\hspace{0.5cm} \gop = \frac{4 G^2}{\kappa},
\end{align}
with $G=g_0\bar{a}_p$ being the many-photon optomechanical coupling rate.

Our interest is on the output field leaving the right port of the cavity, and hence on the first row of $\mathbf{s}^{\pm}[\omega]$.  For a weak optomechanical cooperativity, we can ignore
the modification of the mechanical damping by the cavity, and approximate $\gamma_m \pm \gop \simeq \gamma_m$.
The only remaining differences in the first row of $\mathbf{s}^{+}$ versus $\mathbf{s}^{-}$ are in the overall sign of the mechanical contributions (terms $\propto \gop$) in the elements $\mathbf{s}_{11}$ and $\mathbf{s}_{12}$.  These elements describe how the incident microwave fluctuations show up in the output; the sign difference of the mechanical term directly mirrors the fact that for the red (blue) detuned drive, the cavity provides positive (negative) optical damping on the mechanics.  Note finally that for weak coupling, 
 the coefficient $\mathbf{s}_{13}$ describing the transmission of mechanical bath fluctuations to the output is identical for both choices of drive detuning.

The normal ordered noise spectral density and the symmetrized noise spectral density of the output field on the right side of the cavity are defined as
\begin{align} \label{Eq.DefinitionsSpectra}
\SNII[\omega] =& \int dt\left\langle :\hat{I}\left(t\right)\hat{I}\left(0\right):\right\rangle e^{i\omega t}, 
\hspace{0.5cm}
\SbII[\omega] = \frac{1}{2}\int dt\left\langle \left\{ \hat{I}\left(t\right),\hat{I}\left(0\right)\right\} \right\rangle e^{i\omega t},
\end{align}
where $\hat{I}\left(t\right)=\hat{d}_{R,\rm out}^{\PD}(t)+\hat{d}_{R,\rm out}^{\dagger}(t)$ is defined in terms of {\it lab frame output operators}.  This definition makes the symmetrized
noise spectral density consistent with that measured by a classical voltage spectrum analyzer, as used in the experiment.  Note that to describe a homodyne measurement, one should instead take
$\hat{I}$ to be defined in terms of output operators {\it in the rotating frame}.  This difference only affects the frequency-independent noise floor.  In our case, we will focus on frequencies near the cavity resonance frequency (i.e.~$\omega \simeq \omega_c$ in the lab frame).  For such frequencies, terms in the spectra involving the output operator $\hat{d}^{\dag}_{R,\rm out}(t)$ will not contribute, as these operators
only have spectral weight at negative frequencies in the lab frame (see, e.g.,  Appendix D in \cite{Clerk:2010}.)  We can thus replace 
$\{ \hat{I}\left(t\right),\hat{I}\left(0\right)\} $ 
by
$\{ \hat{d}_{R,\rm out}^{\PD}(t) , \hat{d}_{R,\rm out}^{\dagger}(0) \}$ in the definition of the symmetrized spectrum.  Correspondingly, we can replace the expectation value
in the definition of the normal ordered spectrum by $\langle \hat{d}_{R,\rm out}^{\dag}(0) \hat{d}_{R,\rm out}^{\PD}(t) \rangle$.

Having established the definition of the noise spectra, we now return to our rotating frame, where the cavity frequency is situated at $\omega = \Delta$.
By using the correlators defined in Eq.~(\ref{Eq.correlators}), we can calculate these noise spectral densities and
express them in terms of the elements of the scattering matrix Eq.~(\ref{Eq.ScatteringMatrix}). 
We obtain for the symmetrized spectra:
\begin{align}
\SbII[\omega] \Big|_{\Delta = \pm \omega_m} &=	  
					    | s_{11}^{\pm}[\omega]|^2 \left( \nRth + \frac{ \alpha_{R}}{2}\right) 
					 +  | s_{12}^{\pm}[\omega]|^2 \left( \nLth + \frac{ \alpha_{L}}{2}\right) 
		           		 +  | s_{13}^{\pm}[\omega]|^2 \left( \nmth + \frac{\beta}{2} \right),
\end{align}
while the normal-ordered spectra take the form: 
\bse  
\begin{align}
\SNII[\omega] \Big|_{\Delta = \omega_m} &=	  
					    | s_{11}^{+}[\omega]|^2  \nRth
					 +  | s_{12}^{+}[\omega]|^2  \nLth
		           		 +  | s_{13}^{+}[\omega]|^2  \nmth,
\\
\SNII[\omega] \Big|_{\Delta = - \omega_m} &=	  
					    | s_{11}^{-}[\omega]|^2  \nRth
					 +  | s_{12}^{-}[\omega]|^2  \nLth 
		           		 +  | s_{13}^{-}[\omega]|^2  \left( \nmth + \beta \right).
\end{align}
\ese
Note crucially that for a given drive detuning, {\it the scattering matrix elements appear identically in both the symmetrized and normal-ordered spectra}.  
The only difference is how these elements are weighted by the input noise.  For the symmetrized spectra, it is always the symmetrized bath noise which enters (i.e.,~ $n^{\rm th}_\sigma + 1/2$), irrespective
of the drive detuning.  In the normal ordered case, we see that the only contribution from vacuum noise is from the mechanical bath, and only for the case of a blue-detuned drive.  We also note that the form of the symmetrized spectra given above could be obtained from a completely classical set of Langevin equations, as the input noise correlators enter the same way for both detunings.  This is not true for the normal ordered case, as the effective mechanical bath correlator is different for $\Delta = \omega_m$ versus $\Delta = - \omega_m$.

Setting $ \alpha_{R} = \alpha_{L} \equiv \alpha$ for clarity, the imbalance of the spectra (i.e., the difference between the output spectra for the two choices of detuning) 
$\delta S = S |_{\Delta = - \omega_m} \hspace{-0.2cm} - S |_{\Delta = \omega_m}$, 
become 
\begin{align}
\delta \SbII &=	  
			    \left\{ | s_{11}^{-}|^2 - | s_{11}^{+}|^2  \right\}  \left( \nRth + \frac{ \alpha}{2}\right) 
			 +  \left\{ | s_{12}^{-}|^2 - | s_{12}^{+}|^2  \right\}  \left( \nLth + \frac{ \alpha}{2}\right) 
	           	 +  \left\{ | s_{13}^{-}|^2 - | s_{13}^{+}|^2  \right\}  \left( \nmth + \frac{\beta}{2} \right),
	           	 \nonumber \\ 
\delta \SNII &=	  
			    \left\{ | s_{11}^{-}|^2 - | s_{11}^{+}|^2  \right\}   \nRth  
			 +  \left\{ | s_{12}^{-}|^2 - | s_{12}^{+}|^2  \right\}   \nLth
	           	 +  \left\{ | s_{13}^{-}|^2 - | s_{13}^{+}|^2  \right\}   \nmth
	           	 +  | s_{13}^{-}|^2 \beta.
\end{align}
We have omitted writing the explicit frequency dependence of the elements of $\mathbf{s}^\pm$ for clarity. 

Finally, we insert the explicit elements of the scattering matrix in Eq.~(\ref{Eq.ScatteringMatrix}) into the expressions for the different output spectra derived above. The symmetrized noise in the rotated frame becomes
\begin{align}
\SbII [\omega]\Big|_{\Delta =  \omega_m} \hspace{-0.5cm} =& 
			    \ \bSo
		            +  \frac{\kappa_{ R}}{\kappa}  \frac{\gamma_m \gop}{ (\omega - \omega_m)^2 + \frac{\gtot^2}{4}}
			      \bigg[ \nmth - \neff  +  \frac{\beta - \alpha_{R}}{2}  
			       - \frac{\gop}{\gamma_m} \hspace{-0.12cm} 
			       \left[   \ncth   -  \nRth  \right] \hspace{-0.12cm} 
			       -  \frac{\kappa_{\rm L} }{\kappa}  \hspace{-0.12cm} 
				\left[2 + \frac{\gop}{\gamma_m}\right] \hspace{-0.12cm} 
			       \left(  \hspace{-0.12cm} \frac{ \alpha_{ L} -  \alpha_{R}}{2}  \hspace{-0.12cm}  \right) \hspace{-0.1cm}  \bigg],
\nonumber \\
\SbII [\omega]\Big|_{\Delta = - \omega_m} \hspace{-0.72cm} =& 
			    \ \bSo
		            +  \frac{\kappa_{ R}}{\kappa}  \frac{\gamma_m \gop}{ (\omega + \omega_m)^2 + \frac{\gtot^2}{4}}
			      \bigg[ \nmth + \neff  +  \frac{\beta + \alpha_{R}}{2}  
			       - \frac{\gop}{\gamma_m} \hspace{-0.12cm} 
			       \left[   \ncth   -  \nRth  \right] \hspace{-0.12cm} 
			       +  \frac{\kappa_{\rm L} }{\kappa}  \hspace{-0.12cm} 
				\left[2 - \frac{\gop}{\gamma_m}\right] \hspace{-0.12cm} 
			       \left(  \hspace{-0.12cm} \frac{ \alpha_{ L} -  \alpha_{R}}{2}  \hspace{-0.12cm}  \right) \hspace{-0.1cm}  \bigg],
 \end{align}
where we define $\neff = 2 \ncth - \nRth$, $\gtot = \gamma_m \pm \gop$ and the noise floor
\begin{align}\label{Eq.Noisefloor}
 \bSo =   \frac{\alpha_{ R}}{2} +  \nRth  
		+ \frac{4\kappa_{R}}{\kappa}  \left( \ncth -  \nRth \right)
		+ \frac{2\kappa_{R}}{\kappa}              \left(  \alpha_{L} -\alpha_{R}   \right). 
\end{align}
For the normal-ordered spectra, we obtain:  
\begin{align}
\SNII [\omega]\Big|_{\Delta =  \omega_m} \hspace{-0.5cm} =& 
			    \ \bSo - \frac{\alpha_R}{2}
		            +  \frac{\kappa_{ R}}{\kappa}  \frac{\gamma_m \gop}{ (\omega - \omega_m)^2 + \frac{\gtot^2}{4}}
			      \bigg[ \nmth - \neff   
			       - \frac{\gop}{\gamma_m}  
			       \left[   \ncth   -  \nRth  \right]   \bigg],
 \\
\SNII [\omega]\Big|_{\Delta = - \omega_m} \hspace{-0.72cm} =& 
			    \ \bSo - \frac{\alpha_R}{2}
		            +  \frac{\kappa_{ R}}{\kappa}  \frac{\gamma_m \gop}{ (\omega + \omega_m)^2 + \frac{\gtot^2}{4}}
			      \bigg[ \nmth + \neff  +  \beta  
			       - \frac{\gop}{\gamma_m} 
			       \left[   \ncth   -  \nRth  \right]  \bigg].
 \end{align}
For a weak optomechanical cooperativity, $\gtot \simeq  \gamma_m$.  If assume this case, take $\alpha_L = \alpha_R$, and transform back into the lab frame, 
we recover the spectral densities given in the main text, cf., Eq.~(2,3,5,6).

It is also useful to characterize the asymmetry of the $\Delta = \pm \omega_m$ spectra in terms of the total integrated weight of the mechanical feature.  Defining $\delta \mathcal I = \int \frac{d\omega}{2 \pi} \delta S[\omega]$ and taking $\gop \ll \gamma_m$, we find 
\begin{align}
\delta \bar{\mathcal I}_{II,\rm tot} &=	  
			       \frac{\kappa_{ R}}{\kappa}   \gop \ 
			      \bigg[ 2 \neff  +  \frac{\kappa_{R} }{\kappa} \alpha_{R}  
			       +  \frac{\kappa_{ L} }{\kappa} \alpha_{ L}   \bigg],
\hspace{0.5cm}
\delta \mathcal I_{II,\rm tot}^{\rm N} =	  
				\frac{\kappa_{ R}}{\kappa} \gop \
			      \bigg[ 2 \neff  +  \beta   \bigg].
\end{align}
We thus see that the asymmetry of the symmetrized spectra (corresponding to linear field measurement) are most naturally interpreted as being due to the contribution of fluctuations of
the incident microwave fields, whereas the asymmetry in the normal ordered spectra are most naturally attributed to the fluctuations of the mechanical oscillator.

Finally, note that the output spectra are linked via the commutation relation of the output field, which must be the same as those of the corresponding input field:
$$\left[\hat{d}_{R,\rm out}^{\PD} [\omega],\,\hat{d}_{R,\rm out}^{\dagger} [\Omega]\right]=\alpha_{R} \ \delta(\omega + \Omega).$$ 
Calculating the commutator using the scattering matrix in Eq.~(\ref{Eq.ScatteringMatrix}) and keeping $\alpha_L, \alpha_R$ and $\beta$ unspecified, we obtain for both detuning cases 
\begin{align}
 \left[ \hat d_{ R,\rm out}^{\PD}[\omega], \hat d_{R,\rm out}^{\dag}[\Omega]  \right] \Big|_{\Delta = \pm\omega_m} 
			  \hspace{-0.7cm}  =&  \ \
			 \delta(\omega + \Omega) \ \bigg\{   \alpha_{R} 
					      +\frac{4\kappa_{ R}\kappa_{L} }{\kappa^2}   \left(\alpha_{L}  - \alpha_{R}   \right)
					      \nonumber \\ &
				              \pm  \frac{\kappa_{R}}{\kappa}  
				               \frac{  \gamma_m \gop}
				                    { (\omega \mp \omega_m)^2 + \frac{\gtot^2}{4} }
				              \left[ \left(\beta - \frac{\kappa_{L}}{\kappa} \alpha_{L}  - \frac{\kappa_{R}}{\kappa} \alpha_{R}  \right)
				                \mp \frac{\kappa_{L}}{\kappa}
				                \left( 1 \pm \frac{\gop}{\gamma_m}  \right)
				                \left(   \alpha_{R} - \alpha_{L} \right) \right]
				               \bigg\}.
\end{align}
We thus see that preserving the commutation relation of the output $R$ fields requires in general $\alpha_L = \alpha_R = \beta$.  The fact that the commutator of the output field is a constant means that for any detuning, the symmetrized spectrum will be equal to the normal ordered spectrum plus a frequency-independent noise background.


\subsection*{Balanced Detuned Two Tones with Cooling}
In our actual experiment, we have a two-port electro-mechanical system, which we pump during our measurement simultaneously with three microwave tones.
These tones are all detuned from the cavity resonance, and in a frame rotated at the cavity frequency $\omega_c$ the drive Hamiltonian reads
\begin{align} \label{Eq.TTdrive}
 \hH_{\rm drive} =  \sum_{\nu= {\mp}}   a_{\nu}  \left( \hat a \ e^{i \nu \left( \omega_{m} + \delta \right) t} 
						+  \hat a^{\dag} e^{-i \nu \left( \omega_{m} + \delta \right)t} \right) 
					+  a_{\rm cool}  \left( \hat a \ e^{-i \left( \omega_{m} + \delta_{c} \right) t} 
						+  \hat a^{\dag} e^{ i \left( \omega_{m} + \delta_c \right)t} \right),
\end{align}
the first term describes the balanced detuned two tones: one is in the amount of $\delta$ detuned below the red sideband ($\nu = -$) and the other one is with the same amount detuned above the blue sideband ($\nu = +$). The second term corresponds to the cooling tone, which we assume to be sufficiently detuned below the red sideband ($\delta_c > \delta \gg \gamma_m$), so that the cooling tone acts independently from the probe tones. 

Now, we start with the driving scheme in Eq.(\ref{Eq.TTdrive}) and the standard Hamiltonian in Eq.~(\ref{Eq.StandHamil}), which we rotate in a frame at the cavity frequency and the mechanical frequency $\omega_m$. Additionally, we perform a rotating wave approximation as usual, where we neglect non-resonant processes ($\omega_m\gg \kappa$). 
As before we use input-output theory to include the dissipative environment and derive the quantum Langevin equations for the fluctuation operators of the microwave (mechanical) $\hat d (\hat c)$ system.
By solving these Langevin equations for the noise operator $\hat{c} [\omega]$ of the mechanical
oscillator, we can derive the symmetrized noise spectral density of the mechanical motion 
$( \hat{x}=\left[\hat{c}+\hat{c}^{\dagger}\right] \xzp)$
\begin{align} \label{Eq.Sxx}
\bar{S}_{xx} [\omega] & = 
				    \frac{1}{2} \int dt \ e^{i\omega t}  \left\langle \left\{ \hat{x}\left(t\right),\hat{x}\left(0\right)\right\} \right\rangle
				  = \frac{\gamma_M}{\omega^2 + \frac{\gtot^2}{4} } 
					\left[ \left( \nMth + \frac{1}{2}\right)
					      + \frac{\gop^{-} + \gop^{+} }{\gM} \left( \ncth + \frac{1}{2}\right)\right]
				    \xzp^{2},
\end{align}
with the total damping $\gtot = \gM + \gop^{+} - \gop^{-}$, where the $\gop^{\pm} = 4 G_{\pm}^2/\kappa$ corresponds to the optical damping/antidamping induced by the red/blue tone. The optical damping $\gop^{\rm cool}$, associated with the cooling tone at $\omega = \omega_c - \omega_m -\delta_c$, is included in the enhanced mechanical linewidth $\gM = \gamma_{m} + \gop^{\rm cool}$, as well as in the modified mechanical occupation   $\nMth = ( \gamma_m\nmth + \gop^{\rm cool}\ncth )/\gM $.

In the calculation of the output spectra we assume that the anti-Stokes sideband created by the red tone and the Stokes sideband created by the blue tone, can be treated independently. The distance between the two sideband in frequency space is $2\delta$, thus for $\delta \gg \gtot$ we have two well separated Lorentzians and we can neglect a direct coupling of the drives in the Langevin equations; see the next section for further discussions.
The noise in the output field near the Stokes sideband ($\omega = \omega + \delta$) and  anti-Stokes sideband ($\omega = \omega - \delta$) becomes
\begin{align} \label{Eq.OutputOperator}
\hat d_{R,\rm out}[\omega \mp \delta]  &=  
                  \left[ 1 -    \frac{2\kappa_{R}}{\kappa}
		           \pm  \frac{ \kappa_{R}}{\kappa}
	          		\frac{  \gop^{\pm}}
                  		     { -i \omega + \frac{\gtot}{2} }   \right]
		   \hat d_{R,\rm in}[\omega \mp \delta]
                 -\left[ 2 \mp  \frac{ \gop^{\pm}}
                                   {  -i \omega + \frac{\gtot}{2}} \right] 
		   \sum_{\bar{\sigma} \in {L,I}} \frac{\sqrt{\kappa_{R}\kappa_{\bar \sigma}}}{\kappa} \ \hat d_{\bar{\sigma}, \rm in}[\omega \mp \delta] 
\nonumber \\  &   
		    + i \sqrt{\frac{\kappa_{\rm R}}{\kappa}}
			\frac{ \sqrt{\gM  \gop^{\pm} }}
			  { -i \omega + \frac{\gtot}{2}} 
		     \  \hat c_{\rm in, tot}^{(\dag)}[\omega]
                   \pm  \frac{ \sqrt{\gop^{-} \gop^{+}}}
			     { -i \omega + \frac{\gtot}{2}}  
		    \sum_{\sigma \in {L,I,R}} 
			\frac{\sqrt{\kappa_{ R}\kappa_\sigma}}{\kappa} 
			      \hat d_{\sigma,\rm in}^{\dag}[\omega \mp \delta] ,
\end{align}
with the effective mechanical input noise
\begin{align}
 \hat c_{\rm in,tot}^{(\dag)} [\omega ] = \sqrt{\frac{\gamma_m}{\gM}} \ \hat c_{\rm in}^{(\dag)}[\omega] 
				\mp i \sqrt{\frac{\gop^{\rm cool}}{\gM}} \sum_{\sigma \in {L,I,R}} \sqrt{\frac{ \kappa_\sigma}{\kappa}} \ 
				\hat d_{\sigma,\rm in}^{(\dag)} [\omega \mp \delta_c],
\end{align}
where we again used the input-output relation $\hat d_{R, \rm out} = \hat d_{R, \rm in} + \sqrt{\kappa_R} \hat d$ and approximated the susceptibility of the cavity as $\chi_c [\omega] = [-i(\omega \mp \delta_{(c)}) + \kappa/2]^{-1} \simeq 2/\kappa$.
The structure of the output noise operator is similar to the single tone setup, cf. first row in Eq.~(\ref{Eq.ScatteringMatrix}). Though here we have also a contribution from the cooling tone and a coupling to $\hat d_{\sigma,\rm in}^{\dag}$, arising from the fact that the mechanical oscillator sees both drives and thus, mediates an indirect coupling between the two sidebands.
With the noise correlators and commutation relations given in Eq.~(\ref{Eq.correlators}) and setting $\alpha_\sigma = \beta = 1$, the symmetrized noise spectral densities are
\bse
\begin{align}
\SbII[\omega - \delta] &= \bSo
			+ \frac{\kappa_{R}}{\kappa} \frac{ \gtot \gop^{+} }
			  { \omega^2 + \frac{\gtot^2}{4} } 
			 \left[ \nm - \neff \right], \label{Eq.SymSpectraRed}
			      \\
\SbII[\omega + \delta] &= \bSo
			+ \frac{\kappa_{R}}{\kappa} \frac{  \gtot  \gop^{-} }
			  { \omega^2 + \frac{\gtot^2}{4} } 
			 \left[ \nm  + \neff +     1 \right], \label{Eq.SymSpectraBlue}
\end{align}
\ese
with the averaged mechanical occupation
\begin{align}
\nm =& \frac{\gM}{\gtot} \nMth
		+ \frac{\gop^{-}}{\gtot} \left(  \ncth + 1\right) + \frac{\gop^{+}}{\gtot} \ncth  ,
\end{align}
and the noise floor $\bSo$ as defined in Eq.~(\ref{Eq.Noisefloor}). 
The normal ordered noise spectral densities yield
\bse
\begin{align}
 \SNII [\omega - \delta] &= \bSo - \frac{1}{2}
			+ \frac{\kappa_{R}}{\kappa} \frac{  \gtot  \gop^{+} }
			  { \omega^2 + \frac{\gtot^2}{4} } 
			 \left[ \nm  - \neff \right],
			      \\
 \SNII [\omega + \delta] &= \bSo - \frac{1}{2}
			+ \frac{\kappa_{R}}{\kappa} \frac{  \gtot  \gop^{-} }
			  { \omega^2 + \frac{\gtot^2}{4} } 
			 \left[ \nm +  \neff + \frac{\gM}{\gtot} + \frac{\gop^{+} - \gop^{-}}{\gtot}    \right].
\end{align}
\ese
For equal coupling strengths $G_{+}=G_{-}$ the amount of optical damping and antidamping is the same, i.e., $\gop^{+} = \gop^{-} = \gop$, and thus the total damping contains only the enhanced mechanical linewidth $\gtot = \gM$. For the case $G_{+} \neq G_{-}$ the asymmetries in terms of their integrated weights become
\bse
\begin{align}
 \delta \bar{\mathcal I}_{II,\rm tot} = \delta \mathcal I^{\rm N}_{II,\rm tot} =& \frac{\kappa_R}{\kappa} \left[
				   \nm \left[ \gop^{-} - \gop^{+} \right]
				+  \left(  \neff + 1 \right) \gop^{-} 
					+  \neff \gop^{+}   \right] .
\end{align}
\ese
Thus, the observed sideband asymmetry of the
symmetrized spectrum  and the sideband asymmetry for the normal order coincide. For balanced optical damping rates $\gop^{+}= \gop^{-}$, we obtain the expected result for the asymmetry, which scales with $2 \neff + 1$.


\subsection*{Effects of asymmetric parameters and next sideband contributions}
\begin{figure}[t] 
\centering
 \includegraphics[width=0.99\textwidth]{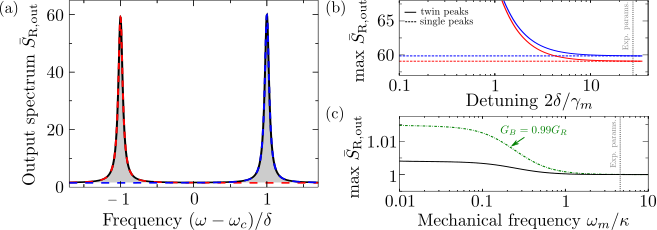}
 		\caption{(a) Symmetrized output spectra as a function of frequency, for experimental parameters. The black solid line shows the non-RWA result, 
			which exhibits two peaks at $\omega = \omega_c \pm \delta$. 
			The dashed red and dashed blue line correspond to two independent noise spectral densities, one valid for frequencies close to the anti-Stokes sideband (Eq.~(\ref{Eq.SymSpectraRed})), and the other one for frequencies close to the Stokes sideband (Eq.~(\ref{Eq.SymSpectraBlue})),
			which each accurately describes the resonant peaks at $\omega - \omega_c = \mp \delta$ .
			(b) Influence of the detuning parameter $\delta$ on the output noise. The solid red (blue) line corresponds to noise maximum at $\omega - \omega_c = - \delta$ ($\delta$) calculated with the twin peaks spectrum, i.e., from Eq.~(\ref{Eq.CompleteSpectrum}). 
			For $\delta \gg \gtot = \gM$ these results coincide with the single peak approximations  (dashed lines). (c) Effect of next sideband contributions on the noise maximum. The dashed-dotted line includes a $1 \% $ mismatch of the coupling strengths.
			Experimental parameters: $\omega_m/(2 \pi) = 4$~MHz, $\gM/(2 \pi) = 360$~Hz , $\delta/(2 \pi) =  5 $~kHz,  $\kappa/(2 \pi) = 870$~kHz, $\kappa_{ R}/(2 \pi) = 450$~kHz, $\kappa_{L}/(2 \pi) = 155$~kHz. For the cavity baths we assumed $\nRth = \nLth = 0.3, \ncth = 0.24 $ and for the mechanical bath  $\nMth = 100$. \label{Corrections} }
\end{figure}
In our main analysis, we assumed that the direct coupling between the cavity fields at the two mechanically-generated sidebands near the cavity resonance is negligible.
By this, we mean that the Lorentzian-shaped resonance around $\omega_c - \delta$ (lab frame), created due to the drive on the red sideband,
does not overlap with the one created from the drive on the blue sideband at frequency $\omega_c +\delta$. 
Within this approximation we could derive two independent noise spectral densities, one valid for frequencies close to the Stokes sideband Eq.~(\ref{Eq.SymSpectraBlue}), and the other for frequencies close to the anti-Stokes sideband Eq.~(\ref{Eq.SymSpectraRed}). From these expressions follows, that the width of each Lorentzian is given by the total damping rate $ \gtot$. Thus, the detuning of the control lasers from the sideband frequencies should fulfill the condition $\delta \gg \gtot $. 
Briefly, we want to confirm the validity of this condition by calculating the complete RWA solution.
In this case, the noise in the output field near the cavity resonance becomes ($G_{-}=G_{+}$ and rotated frame)
\begin{align}
 \hat d_{R, \rm out}^{\PD}[\omega ] = \ \hat d_{R,\rm in}^{\PD}[\omega ]
			&-  \sqrt{\kappa_{R}}  \chi_c[\omega ] 
			    \sum_{\sigma\in R,L,I} \sqrt{\kappa_\sigma }  \hat d_{\sigma,\rm in}^{\PD}[\omega ] 
\nonumber \\ &
			-  \sqrt{\frac{\kappa_{R}}{\kappa}} \gop \left(\chi_m [\omega - \delta] - \chi_m [\omega +  \delta] \right)
			 \sum_{\sigma\in R,L,I} \sqrt{\frac{\kappa_\sigma}{\kappa}} \left( \hat d_{\sigma,\rm in}^{\PD}[\omega ]  +\hat d_{\sigma,\rm in}^{\dag}[\omega ]  \right)
\nonumber \\ & 
			+ i  \sqrt{ \frac{\kappa_{R}}{\kappa} \gM \gop} 
			\left( \chi_m [\omega+\delta]    \ \hat c_{\rm in,tot}^{\PD}[\omega+\delta] 
			     + \chi_m [\omega -\delta ]  \ \hat c_{\rm in,tot}^{\dag}[\omega - \delta ] \right),
\end{align}
with the susceptibilities for the mechanical oscillator $\chi_m[\omega] = [-i \omega + \frac{\gM}{2} ]^{-1}$ and the microwave cavity $\chi_c[\omega] = [-i \omega + \frac{\kappa}{2} ]^{-1}$. Note, that near the Stokes/anti-Stokes sideband, i.e., for frequencies around $\omega \sim  \pm \delta$, the mechanical susceptibility shows up as $\chi_m [\omega]$ and $\chi_m [\omega \pm 2 \delta]$, by neglecting the contribution from $\chi_m [\omega \pm 2 \delta]$ we recover Eq.~(\ref{Eq.OutputOperator}),  describing two independent resonances at $\omega \sim \pm \delta$.

\noindent
For simplicity, we focus on the symmetrized noise spectral density, which in this case yields
\begin{align} \label{Eq.CompleteSpectrum}
 \SbII[\omega ] =& \ \bSo
			- \frac{4 \kappa_{R}}{\kappa}  
			  \frac{ \gop^2 \left[ (\omega - \delta)(\omega+ \delta) + \frac{\gM^2}{4}\right]}
				{\left[ (\omega+ \delta)^2 + \frac{\gM^2}{4}\right] \left[ (\omega - \delta)^2 + \frac{\gM^2}{4}\right]}
			   \left(\ncth + \frac{1}{2}\right) 
\nonumber \\ & 
			+  \frac{\kappa_{R}}{\kappa} 
			  \frac{\gM \gop}{(\omega+ \delta)^2 + \frac{\gM^2}{4} }   
			  \left(\nm    -  \neff  \right)
			+  \frac{\kappa_{R}}{\kappa} 
			   \frac{\gM \gop}{(\omega - \delta)^2 + \frac{\gM^2}{4} } 
			  \left( \nm +  \neff + 1\right)
\nonumber \\
			=& \ \bSo + \SbII^{\rm mix}[\omega]
			 + \SbII^{\rm AS}[\omega ]
			 + \SbII^{\rm S}[\omega].
\end{align}
Here, we have written the noise spectral density in a frame rotating at the cavity resonance frequency.
This expression contains both Lorentzian near the anti-Stokes (AS) and Stokes (S) sideband, as well as the noise floor $\bSo$ 
and a mixing term $\SbII^{\rm mix}[\omega]$. Figure \ref{Corrections}(a) shows a plot of this output spectrum for the parameters used in the experiment. Both resonances are clearly separated and each well described by the spectra calculated without a coupling of the fields (dashed red/blue lines).
By decreasing the detuning $\delta$ the distance between the peaks decreases and they start to overlap. 
Without any detuning we end up with a Lorentzian at the cavity resonance, with an integrated weight
containing solely the mechanical bath. 

To study the influence of the detunings, we compare the symmetrized output spectrum Eq.~(\ref{Eq.CompleteSpectrum}) 
to the single Lorentzian approximations $\SbII^{\rm{S/AS}}[\omega]$. For simplicity we focus on the maxima at the Stokes and anti-Stokes sideband and subtract the noise floor. 
We end up with the ratios
\begin{align}
 \frac{\SbII[\pm \delta] - \bSo }{\SbII^{\rm{S/AS}}[\pm \delta]}
			=& \ 1 + 
			  \frac{1}{\left[ \frac{4\delta}{\gM} \right]^2 + 1 }   
			  \left( \frac{ \nMth  - n_{\rm opt}^{\rm th}  + \frac{1\mp 1}{2} }{\nMth + n_{\rm opt}^{\rm th}  + \frac{1 \pm 1}{2} }
			  \right),
\end{align}
with $ n_{\rm opt}^{\rm th} = \frac{\gop}{\gM} \left( 2 \ncth + 1 \right)  \pm \neff$. 
From this we see, that the corrections to $\SbII^{\rm{S/AS}}[\omega]$ (second term in the equation above) vanish as expected, if $\delta \gg \gM$, cf. Fig. \ref{Corrections}(b). Hence, in this regime we can describe our spectra as two individual resonances. 
Note, that the actual chosen detuning in the experiment lays clearly in the regime where both peaks are well separated, cf. Fig. \ref{Corrections}(a,b).

Finally, we want to briefly comment on the influence of higher-order mechanical sidebands. The general linearized interaction Hamiltonian for our setup reads
\begin{align}
 \hat{\mathcal{H}}_{\rm int} &= 
			 G_{+} 
			\left[ e^{-i \delta t} \hat d \hat c^{\dag}
			    +  e^{i  \delta t} \hat d^{\dag} \hat c \right]
		       + G_{-} 
			\left[  e^{i  \delta t} \hat d \hat c
                             +  e^{-i \delta t} \hat d^{\dag} \hat c^{\dag}\right] + \hat{\mathcal{H}}_{\rm CR},
\nonumber \\ 
 \hat{\mathcal{H}}_{\rm CR} &=
			 G_{\rm +} 
			\left[  e^{i (2 \omega_{m} + \delta) t} \hat d^{\dag}  \hat c^{\dag}
			     +  e^{-i (2 \omega_{m} + \delta) t} \hat d \hat c \right]
		       + G_{\rm -} 
			\left[  e^{i (2 \omega_{m} +  \delta) t} \hat d \hat c^{\dag}
                             +  e^{-i (2 \omega_{m} + \delta) t} \hat d^{\dag} \hat c\right],
\end{align}
where the counter-rotating terms in $ \hat{\mathcal{H}}_{\rm CR}$ describe the strongly non-resonant Stokes and anti-Stokes processes 
generated by the two control lasers. The coupling strengths $G_{\pm} $ contain the drive amplitudes as usual, but we assume again that they can be different in magnitude, which leads to a total mechanical damping of $\gop = \gM +\gop^{+} - \gop^{-}$ with $\gop^{\pm} = 4 G_{\pm}/ \kappa$. Note, for the system to be stable the total damping has to be positive, which roughly translates into the condition $G_{+} > G_{-} $. 

The inclusion of the counter-rotating terms leads to a time-dependent problem, which can not be solved exactly. In principle, $ \hat{\mathcal{H}}_{\rm CR}$ generates an infinite number of sidebands at multiples from $\pm \omega_m$. 
If one is not too far from the resolved sideband limit, a perturbative approach keeping track of only the leading-order sidebands created by $\hat{\mathcal{H}}_{\rm CR}$ is sufficient.  Figure \ref{Corrections}(c) depicts the maximum of the symmetrized spectral density as a function of $\omega_m/\kappa$ including the next sidebands at frequencies $\omega = \omega_c \pm 2\omega_m$.
As expected, when one even modestly approaches the resolved sideband regime, i.e., $\omega_m > \kappa$, the contributions from the counter-rotating terms are negligible. Moreover, for the given experimental setup we are far in the resolved side-band regime as indicated in the graph.


\section{Linear Response Theory}

In this section we briefly review the linear response approach to understand the sideband asymmetry
observed using linear field detection; this explanation was first discussed in Ref. \cite{Khalili:2012}.  For linear field detection the
observed asymmetry can be fully attributed to noise correlations in the detector (in this case, the driven cavity), correlations that could exist classically.  We generalize the discussion 
of Ref. \cite{Khalili:2012} to include thermal noise driving the cavity, showing the same backaction-imprecision correlations allow one to understand the
squashing of thermal noise seen in previous experiments \cite{Roucheleau:2010}.  We also show that the particular value of the backaction-imprecision correlator,
required to account for the zero-temperature sideband asymmetry, plays a special role in the linear-response approach to quantum measurements
\cite{Clerk:PRB,Clerk:2010}:  it is {\it precisely} the value needed to ensure there is no additional constraint on the detector's symmetrized noise
correlators besides what would exist classically.

Following Ref.~\cite{Clerk:2010}, the general linear response approach starts by assuming a linear coupling between the detector and the observable to be measured (in this case $\hat{x}$, the
mechanical position):
\begin{equation}
	\hat{H}_{\rm int}=-A \hat{x}\hat{F}.
\end{equation}
Here, $\hat{F}$ is the detector quantity which couples to the measured system, and plays the role of a backaction force.  In our optomechanical case, we have 
$\hat{F} = -(g / \xzp) \hat{a}^\dagger \hat{a}$.  $A$ is a dimensionless coupling constant that we will use to track the order at which $\hat{H}_{\rm int}$ appears in expressions; 
we will set it to one at the end of the calculation.

Next, consider the detector output observable $\hat{I}$.   We assume that this quantity responds linearly to the mechanical position, 
\begin{eqnarray}
	\langle \hat{I}[\omega] \rangle = -A \chi_{IF}[\omega] \langle \hat{x}[\omega] \rangle,
\end{eqnarray}
where $\chi_{IF}[\omega]$ is the response coefficient or ``forward-gain" of the detector; it is given by a standard Kubo formula.  We are interested in understanding the fluctuations of the detector output.  Quantum linear response theory tells us that these can be completely understood within an equivalent classical stochastic model \cite{Clerk:PRB,Clerk:2010}, where we now replace the operators $\hat{I}(t), \hat{F}(t)$ and $\hat{x}(t)$ by classical random variables.  The fluctuations of the output in this model are written:
\begin{eqnarray}
	\delta I_{\rm tot}[\omega] = \delta I[\omega] - A \chi_{IF}[\omega] \left( \delta x_0[\omega] + A \delta x_{\rm BA}[\omega] \right).
\end{eqnarray}
The first term here represents the intrinsic fluctuations of the output in the absence of any coupling to the mechanics (the imprecision noise).  $\delta x_0[\omega]$ describes the position fluctuations of the mechanics in the absence of any backaction, whereas $\delta x_{\rm BA}[\omega]$ describes the additional backaction-driven fluctuations of the mechanical resonator.  $\delta x_0[\omega]$ is due to the intrinsic mechanical dissipation.  Assuming this dissipation to be in thermal equilibrium, these fluctuations are described by the spectral density
\begin{eqnarray}
	\mathcal{S}_{xx,0}[\omega] \equiv \left \langle | \delta x_0[\omega] |^2 \right \rangle = - \hbar \textrm{ Im } \chi_{xx}[\omega] \coth \beta \hbar \omega/2
	\simeq - \hbar \textrm{ Im } \chi_{xx}[\omega] (1 + 2 \nmth),
\end{eqnarray}
where $\chi_{xx}[\omega]$ is the mechanical force susceptibility,
\begin{align}
	\chi_{xx}\left(\omega\right) & =\frac{1/m}{\left(\omega^{2}-\omega_{m}^{2}\right)+i\omega \gamma_m}.
\end{align}
Similarly, the backaction-driven position fluctuations are described by
\begin{eqnarray}
	\mathcal{S}_{xx,\rm BA}[\omega] \equiv \left \langle | \delta x_{\rm BA}[\omega] |^2 \right \rangle   =  \left| \chi_{xx}[\omega] \right|^2 \mathcal{S}_{FF}[\omega].
\end{eqnarray}
The spectral density of the output fluctuations are then given by
\begin{align}
	\label{eq:ClassicalOutputNoise}
	\mathcal{S}_{II,\rm tot} \left[\omega \right] & =
		\mathcal{S}_{II} \left[\omega \right]  + A^2 \left| \chi_{IF}[\omega] \right|^2 
			\left(
				  \mathcal{S}_{xx,0}[\omega] + A^2 \mathcal{S}_{xx,\rm BA}[\omega] -
			  	2 \textrm{Re } \left(
			  		\chi_{xx}[\omega]^* \frac{\mathcal{S}_{IF}[\omega]}{\chi_{IF}[\omega]} \right)
		\right)  \\
	& = 		\mathcal{S}_{II} \left[\omega \right]  + \left| \chi_{IF}[\omega] \right|^2 
			\left( S_{xx,{\rm eff}}[\omega]
		\right)  .
\end{align}
Here, $\mathcal{S}_{IF}[\omega]$ is the spectral density which describes any possible correlations between the intrinsic imprecision noise contribution to the output ($\delta I[\omega]$) and the backaction force noise driving the mechanics ($\delta F[\omega]$).  In the quantum theory, an identical expression to the above holds, except all classical noise spectral densities 
$\mathcal{S}_{AB}[\omega]$ describing the detector are replaced by the corresponding symmetrized quantum noise spectral densities $\bar{S}_{AB}[\omega]$ \cite{Clerk:2010}:
\begin{eqnarray}
	\bar{S}_{AB}[\omega] =
		\frac{1}{2} \int_{-\infty}^{\infty} dt \left \langle \{ \hat{A}(t), \hat{B}(0) \} \right \rangle.
\end{eqnarray}

For our optomechanical system, the needed detector correlation functions are easily computed from the linearized Heisenberg-Langevin equations.  As in the main text, we assume a two-sided cavity, and measure the quantity
$\hat{I}$ defined below Eq.~(\ref{Eq.DefinitionsSpectra}).  One finds
\begin{align}
	\chi_{IF}[\omega] & =  
		- i \frac{ \sqrt{\kappa_R} G}{\xzp } \left(  \chi_c[\omega] - \chi_c[- \omega]^* \right), 
			\\
	\bar{S}_{II} \left[ \omega \right] & =
		\sum_{\nu = \pm}
			\left[         \left| 1 - \kappa_R \chi_c[\nu \omega] \right|^2   \left[ (1/2) + \nRth \right]
			+        \kappa_R \kappa_L  \left| \chi_c[\nu \omega] \right|^2   \left[ (1/2) + \nLth \right] 
		       \right],
			 \\
	\bar{S}_{FF} \left[ \omega \right] & = 
		\frac{G^2}{\xzp^2} \left(
			 \left| \chi_c[\omega] \right|^2 + \left| \chi_c[-\omega] \right|^2 \right)  
			 \sum_{\sigma = L, R} \kappa_\sigma \left[ (1/2) + n^{\rm th}_\sigma \right] ,
			\\
	\bar{S}_{IF} \left[ \omega \right] & =  -\frac{ \sqrt{\kappa_R}  G}{\xzp}
		\sum_{\sigma = L, R} 
			\left(  \Lambda_\sigma[\omega] + \Lambda_\sigma[-\omega]^* \right) 
				\left[ (1/2) + n^{\rm th}_\sigma \right],
\end{align}
where for the cross-correlator, we have introduced the functions
\begin{align}
	\Lambda_R[\omega ] & =
		-  \left( 1 - \kappa_R \chi_c[\omega] \right)  \chi_c[\omega]^* ,
			\\
	\Lambda_L[\omega ] & =
		- \kappa_L \left| \chi_c[\omega] \right|^2 .
\end{align}

We are again interested in the symmetrized output spectrum of the detector in a narrow range ($\sim \gamma_m$) near the cavity resonance frequency, for a drive detuning $\Delta = \pm \omega_m$; as always, we consider the good-cavity limit $\omega_m \gg \kappa$.  Over this range of frequencies, we can neglect the frequency dependence of the cavity correlators, and evaluate them on resonance (i.e., $\omega = \Delta$ in the rotating frame).  Of particular interest is the cross-correlator.  One finds
\begin{align}
	\bS_{zF}[\Delta] \equiv \frac{ \bS_{IF}[\Delta] }{\chi_{IF}[\Delta] }
	& \simeq 
		\mp i \hbar \left[
			\left(  \frac{2 \kappa_R}{\kappa} -1 \right)
			\left[ (1/2) + \nRth \right]  +
			\left(  \frac{2 \kappa_L}{\kappa} \right)
			\left[ (1/2) + \nLth \right]  
		\right]
	\\
	& = 
		\mp i \hbar \left(  \frac{1}{2} + 2 \ncth - \nRth \right),
		\label{eq:SzFFinal}
\end{align}
where the $-$ sign ($+$ sign) corresponds to the drive detuning $\Delta = + \omega_m$ ($\Delta = - \omega_m$).

We see that $\bS_{zF}$ is purely imaginary, and changes sign for the two choices of detuning; in contrast, one can confirm that $| \chi_{IF}|$, $\bS_{II}$ and $\bS_{FF}$ at resonance are the same for $\Delta = \pm \omega_m$.  It immediately follows that the asymmetry between the spectra obtained at $\Delta = -\omega_m$ and $\Delta = \omega_m$ is entirely due to the detector backaction-imprecision correlations described by $\bS_{zF}$.

Returning to Eq.~(\ref{eq:ClassicalOutputNoise}) for the output spectrum, we further note that for a sufficiently weak detector-system coupling, the term $\mathcal{S}_{xx,\rm BA}$ will be negligible to the term $\mathcal{S}_{xx,0}$, as the backaction term is second-order in the coupling (i.e. $\propto A^2$).  However, the last correlation term remains significant:  its contribution relative to $\mathcal{S}_{xx,0}$ is independent of coupling strength.  In our case, where $\mathcal{S}_{IF} / \chi_{IF} \equiv \mathcal{S}_{zF} $ is purely imaginary, we can combine the leading mechanical contributions to the output spectrum as
\begin{eqnarray}
				\mathcal{S}_{xx,{\rm eff}}[\omega] \simeq  
					 \mathcal{S}_{xx,0}[\omega] -
			  	2 \textrm{Re} \left(
			  		\chi_{xx}[\omega]^* \mathcal{S}_{zF} \right) 
 & = & 
	-\textrm{Im} \chi_{xx}[\omega] \left(
		\hbar \coth \beta \hbar \omega/2 + 2 \textrm{Im } \mathcal{S}_{zF}  \right), \\
& \simeq &
	-\textrm{Im} \chi_{xx}[\omega] \left(
		\hbar (1 + 2 \nmth) + 2 \textrm{Im } \mathcal{S}_{zF}  \right). 
	\label{eq:SxxEffFinal}
\end{eqnarray}
We see that the mechanics will give rise to a Lorentzian signature in the output spectrum, but that the presence of imaginary back-action imprecision correlations modifies the weight of the Lorentzian --
it no longer simply reflects the mechanical temperature.  This results in the well known phenomenon of noise squashing.  

Using Eq.~(\ref{eq:SzFFinal}) for the cross-correlator, we see that this linear-response calculation reproduces the asymmetry found earlier between spectra obtained for $\Delta = \pm \omega_m$.  This approach emphasizes the fact that the asymmetry can be completely attributed to the detector, namely the presence of backaction-imprecision correlations.  These correlations are purely imaginary; the only difference between the cases is the sign of the correlator.  For $\Delta = \omega_m$, the correlations are positive, and serve to decrease the weight of the mechanical Lorentzian; they completely cancel the contribution if the mechanics is at zero temperature.  For $\Delta = -\omega_m$, they instead serve to increase the mechanical contribution.  In the absence of thermal cavity noise, the effect of the noise correlations is to cause the weight of the mechanical Lorentzian in the output spectrum to have the expected form for phonon emission or absorption:  for $\Delta = \omega_m$, 
we have the emission factor $\nmth$, for $\Delta = - \omega_m$ we have $\nmth + 1$.

We thus see that the asymmetry can be interpreted in terms of a finely tuned backaction-imprecision noise correlation.  We stress that a completely classical detector could have an identical noise correlation.  Nonetheless, this value of correlation plays an extremely special role in the theory of quantum limits on linear quantum detectors and amplifiers \cite{Clerk:2010}.  Quantum limits on such detectors (e.g.,~on their added noise or noise temperature) follow from a fundamental Heisenberg-like inequality on their noise properties at each frequency.  These take the form:
\begin{eqnarray}
    \bS_{zz}[\omega] \bS_{FF}[\omega] - \left| \bS_{zF}[\omega] \right|^2  & \geq & 
       	 \frac{\hbar^2}{4} 
        \left( 1 +
            \Delta\left[ \frac{2 \bS_{zF}[\omega]}{\hbar  } \right] \right),
        \label{eq:NoiseConstraint}  
\end{eqnarray}
where
\begin{eqnarray}
    \Delta[y] & = &
        \frac{
            \left| 1 + y^2 \right| - \left(1 + |y|^2 \right)
        }{2}.
    \label{eq:DeltaFactor}
\end{eqnarray}
We have defined $\bS_{zz}[\omega] = \bS_{II}[\omega] / | \chi_{IF}[\omega]|^2$, and have specialized to the case relevant here, where the reverse-gain of the detector vanishes.  Note that for any complex number $y$, $1+\Delta[y] > 0$, and hence in general the RHS of the the inequality in Eq.~(\ref{eq:NoiseConstraint}) is non-zero.  This means that in general, quantum mechanics makes it impossible for the detector backaction and imprecision noises to be perfectly correlated (i.e.,~the LHS of Eq.~(\ref{eq:NoiseConstraint}) cannot in general be zero).  This represents a purely quantum constraint on the detector's noise properties.  In the case where $\bS_{zF}$ is purely real (or $0$), the RHS simply reduces to $(\hbar/2)^2$.

In contrast, if $\bS_{zF}$ is purely imaginary, the RHS of Eq.~(\ref{eq:NoiseConstraint}) can be reduced below $\hbar^2 / 4$.  One easily finds that it achieves the minimum value of $0$ when $\bS_{zF} = \pm i \hbar / 2$.  Thus, for this special choice of cross-correlation, quantum mechanics does not forbid a perfect correlation between the detector's backaction and imprecision noises.  
{\it This special value is precisely what was found above for our cavity optomechanical detector at zero temperature} (at the cavity resonance, with a drive detuning $\Delta = \pm \omega_m$, and in the good cavity limit), c.f.~Eq.~(\ref{eq:SzFFinal}).  Thus, not only does this special value of cross correlation yield an imbalance between the $\Delta = \pm \omega_m$ output spectra in just the way expected for quantum emission / absorption, it also implies that there is no additional Heisenberg constraint on the detector noise properties.  As discussed in Sec.~IV.A.4 of Ref.~\cite{Clerk:2010}, this means that in principle, one could make the added noise of this position detector strictly zero.  As is also discussed in this reference, this vanishing does not constitute a violation of the quantum limit on position detection or amplification.  One can demonstrate that in this case, the detector does not provide any amplification of the mechanical motion: the dimensionless power gain of the detector is at most 
order $1$ (see Appendix I.2. of Ref.~\cite{Clerk:2010}).  As there is no amplification, quantum mechanics does not require any added noise.


\section{Device calibrations}

Measurement parameters are deduced from two calibrations. First, we place a single pump tone at ideal ``red'' detuning, $\omega_c -\omega_m$, and monitor the linewidth of the up converted mechanical sideband via weak homodyne detection ($n_p \leq 5\times10^2$).  Sweeping the pump power over $n^{\rm red}_p=10^3-10^7$, we explore the sideband linewidth, $\gtot$, as a function of detected pump power, $P_{\rm thru}^{\rm red}$.  In the resolved sideband regime ($\omega_m \gg \kappa$) with weak coupling ($G \ll \kappa$), the effective linewidth  follows $\gtot=\gamma_m+\frac{4g_0^2}{\kappa}n^{\rm red}_p$.  Fitting to this model, we extract the natural linewidth of the mechanics, $\gamma_m$, and the optical damping as a function of $P_{\rm thru}^{\rm red}$.

Second, two pump tones, denoted as ``+'' and ``-'', are placed at $\omega_\pm=\omega_c\mp(\omega_m+\dP)$, with $\dP=2\pi \times 500$ Hz, and balanced at relatively low powers ($\gamma^-_{op}=\gamma^+_{op}\ll\gamma_m$).  We then measure the integrated noise power of each sideband, $P_m^\pm$, as we sweep the cryostat over calibrated temperatures.  This measurement is performed for both detunings to account for two issues: asymmetric cavity transmission about $\omega_c$ and gain fluctuations at frequencies separated by $\sim 2\omega_m$.  The source of this skewed cavity transmission is addressed in Sec.5.

For small detunings ($\dP \ll \kappa$) and high mechanical occupation factor ($\nmth \gg 1$), the integrated Lorentzian weights of Eqs.~\ref{Eq.SymSpectraRed}, \ref{Eq.SymSpectraBlue} simplify to
\begin{equation}
P_m^\pm= \gain(\omega_c) \cdot \hbar \omega_c \cdot \frac{\kappa_R}{\kappa} \cdot \gop^\pm \cdot n_m^\pm,
\end{equation}
where $\gain(\omega)$ is the system gain between the device output and the room temperature analyzer at frequency $\omega$. We remove the pump power dependence of $\gop^\pm$ by normalizing by the detected tone power, given by $P^\pm_{\rm thru}=\gain(\omega_\pm) \cdot \hbar\omega_\pm \cdot [1+\cavSkew(\omega_\pm)] \cdot \kappa_R \cdot n^\pm_p$.  We include the term $\cavSkew(\omega)$ to incorporate corrections to the microwave transmission mentioned above (See Sec.5).  The resulting ratio is,
\begin{equation}
\frac{P_m^\pm}{P_{\rm thru}^\pm} = \frac{\omega_c}{\omega_\pm} \cdot \frac{\gain(\omega_c)}{\gain(\omega_\pm)} \cdot \frac{1}{1+\cavSkew(\omega_\pm)} \cdot \left(\frac{2g_0 }{\kappa}\right)^2\cdot n^\pm_m.
\end{equation}
For the prescribed cryostat temperatures, the two pump powers are kept low enough ($n^\pm_p\approx10^2$) to ensure that classical noise in the microwave resonator and mechanical bath heating effects are negligible, so that the occupation factor inferred from the sideband areas quantifies the thermal occupation factor of the mechanical mode: $n^\pm_m = \nmth =1/(\rm exp(\frac{\hbar\omega_m}{k_BT})-1)$.  Furthermore, the pump detunings used in the calibration and measurement routines are small enough so that detuning corrections, on the order of $(\frac{2\dP}{\kappa})^2$ can be ignored \cite{Marquardt:2007}.

Combining the above calibrations and extracting the correction factors, $\cavSkew(\omega_\pm)$ (Sec.5), we deduce the optomechanical coupling, $g_0$, which in turn provides the conversion factor between $P^{\pm}_{\rm thru}$ and $n^\pm_p$.  Following these steps, we find that $g_0=2\pi \times 16$ Hz, $n_p^-=1.4\times10^{11} \, \rm W^{-1} \cdot P_{thru}^-$, and $n_p^+=2.6\times10^{11} \, \rm W^{-1} \cdot P_{\rm thru}^+$.


\section{Noise floor calibration}




The increase in the device noise floor at cavity resonance is measured relative to the noise floor of an impedance matched through connection with matching amplifier conditions.  Following Eq.~(\ref{Eq.Noisefloor}) and the noise floor treatment of previous work \cite{Suh:2013}, the noise floor increase is proportional to $\neff$ and $\nRth$,

\begin{equation} \label{Eq.NoisefloorIncrease}
\Delta \eta = \frac{1}{2\etaCF} \left[\neff - \left(\frac{2\kappa_R-\kappa}{2\kappa_R}\right)\nRth \right]
\end{equation} 
where $\neff=2\ncth-\nRth$ as above, and where $\etaCF$ is the conversion factor for $\Delta\eta$ in units of $\ncth$.

To see how this behavior affects our measurements, we consider the sideband powers in the presence of classical noise $\ncth$, $\nRth$.  Integrating the noise power under the Lorentzian of Eqs.~(\ref{Eq.SymSpectraBlue}),~(\ref{Eq.SymSpectraRed}), we find that
\begin{equation}
n_m^+-n_m^- = 4\etaCF\Delta\eta+\left(\frac{2\kappa_R-\kappa}{\kappa_R}\right)\nRth+1,
\end{equation}
and
\begin{equation}
\frac{n_m^++n_m^-}{2} = \left(\frac{2\gop+\gop^{\rm cool}}{\gM}\right) \left[\etaCF \Delta\eta + \left(\frac{4\kappa_R-\kappa}{4\kappa_R}\right)\nRth \right] + \frac{\gamma_m}{\gM}\nmth + \frac{\gop}{\gM} + \frac{1}{2},
\end{equation}
where we follow the notation of Eq.~(\ref{Eq.Sxx}) and have set $\alpha_{\sigma}=\beta=1$, $\gop^+=\gop^-=\gop$.




The $\nRth$ contribution does not affect the slope of either data set.  For sideband imbalance and average measurements, we expect linear dependence on $\Delta\eta$ with slope proportional to $\etaCF$.  The $\nRth$ factor does, however, add fixed offsets to both data sets.  For the sideband difference, the contribution is suppressed relative to the quantum offset of ``+1''.  With the experimental parameters $\nRth=0.34\pm0.03$, $\kappa=2\pi \times (860\pm10)$ kHz, and $\kappa_R=2\pi \times (450\pm30)$ kHz, we estimate an offset correction of $\left(\frac{2\kappa_R-\kappa}{\kappa_R}\right)\nRth\approx(3\pm4) \times 10^{-2}$, well within the measurement uncertainty for sideband imbalance. This is not the case for the sideband average, where we expect a correction to the offset that is significant when compared to the mechanical quantum contribution of ``+1/2''.



\section{Output port occupation}

We estimate the occupation factor of the output port, $\nRth$, by measuring the microwave noise spectrum absent any microwave pumping.  In this setup, we assume that $\ncth$ is solely due to noise radiating into the device from the the isolated port of a cryogenic circulator thermalized to an elevated temperature, so that $\ncth=\nRth\kappa_R/\kappa$. Following \cite{Suh:2013} and Eqs.~(\ref{Eq.Noisefloor}), ~(\ref{Eq.NoisefloorIncrease}), the detected noise floor, now spanning frequencies over the cavity linewidth and also including the noise contribution from the cryogenic amplifier, $\bSamp$, follows
\begin{align}
 \bar{S}_V(\omega) = \frac{1}{\etaCF} \left[
 		\frac{\kappa^2}{\kappa^2+4(\omega-\omega_c)^2}  \left( \frac{\kappa_R}{\kappa} -  1 \right) \nRth
		+ \left(\frac{\kappa}{4\kappa_R}\right) \left(\alpha_{ R} + 2 \nRth \right)
		\right] + \bSamp. 
\end{align}



\begin{figure}[H]
\centering\includegraphics{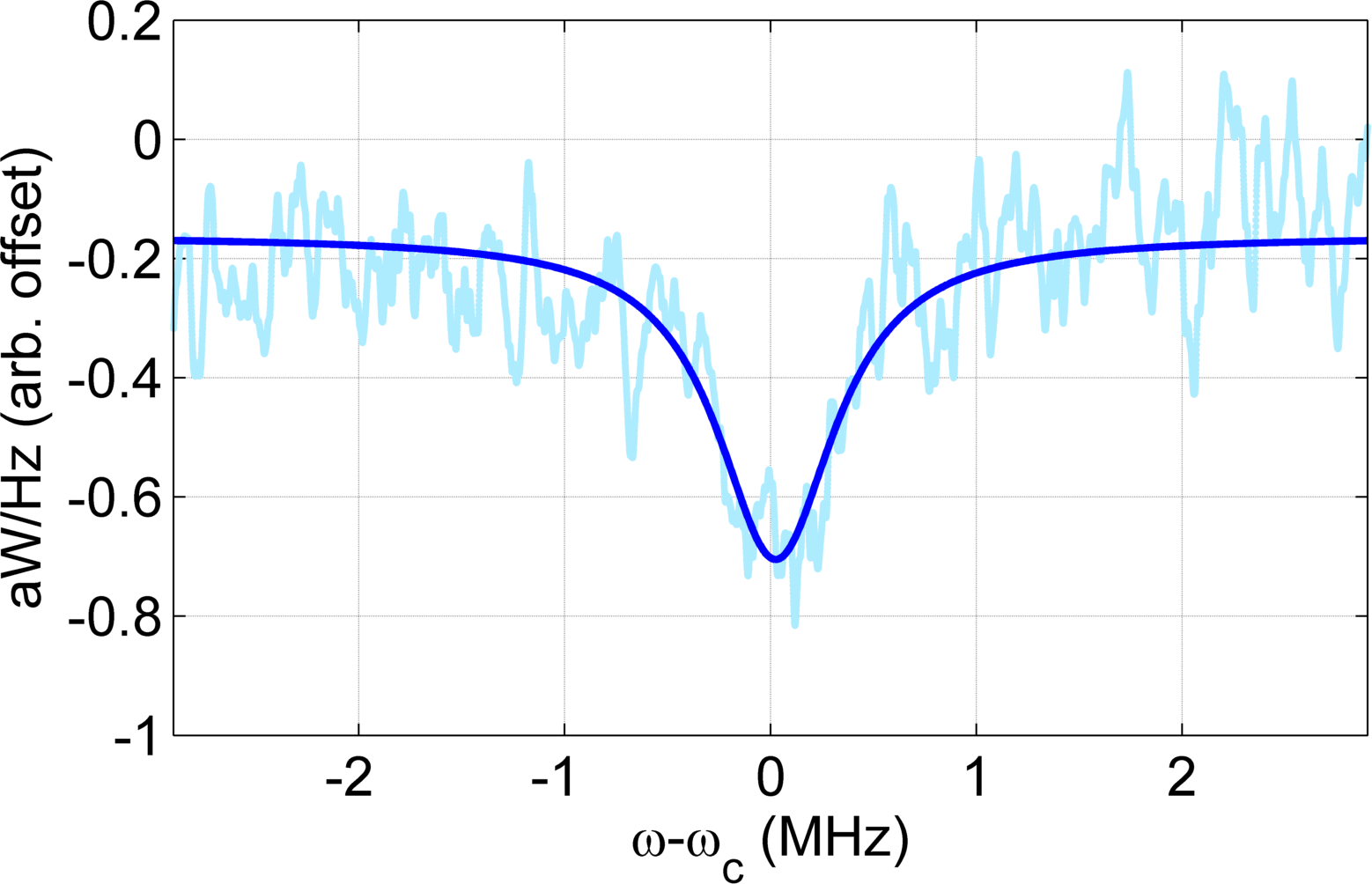}
\caption{Example spectra of microwave noise taken at zero pumping (light blue) with Lorentzian fit (dark blue).} \label{Fig.cavNoise}
\end{figure}

Taking $\frac{\kappa_R}{\kappa}$ from independent calibration measurements and $\etaCF$ from the sideband imbalance measurements, we fit the observed Lorentzian to find $\nRth=(3.4\pm0.3)\times10^{-1}$.  A typical noise floor spectrum is shown in Fig. \ref{Fig.cavNoise}.

\section{Asymmetric microwave transmission}

\begin{figure}[H]
\centering\includegraphics[scale=0.8]{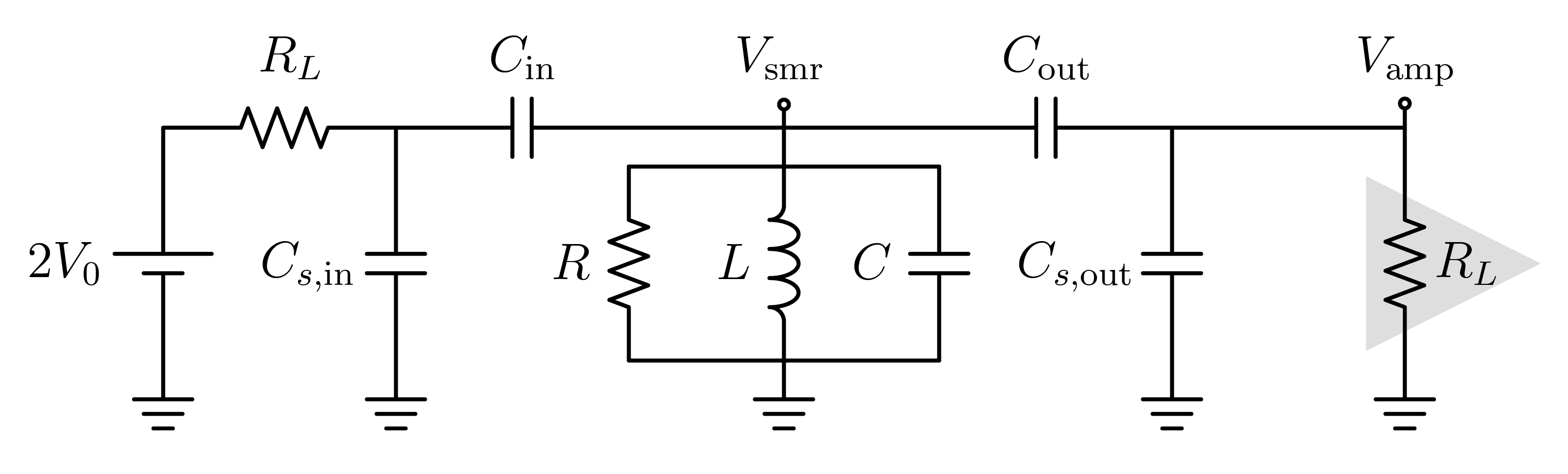}
\caption{Equivalent microwave circuit model with shunt capacitors.} \label{Fig.ShuntCircModel}
\end{figure}

\begin{figure}[H]
\centering\includegraphics{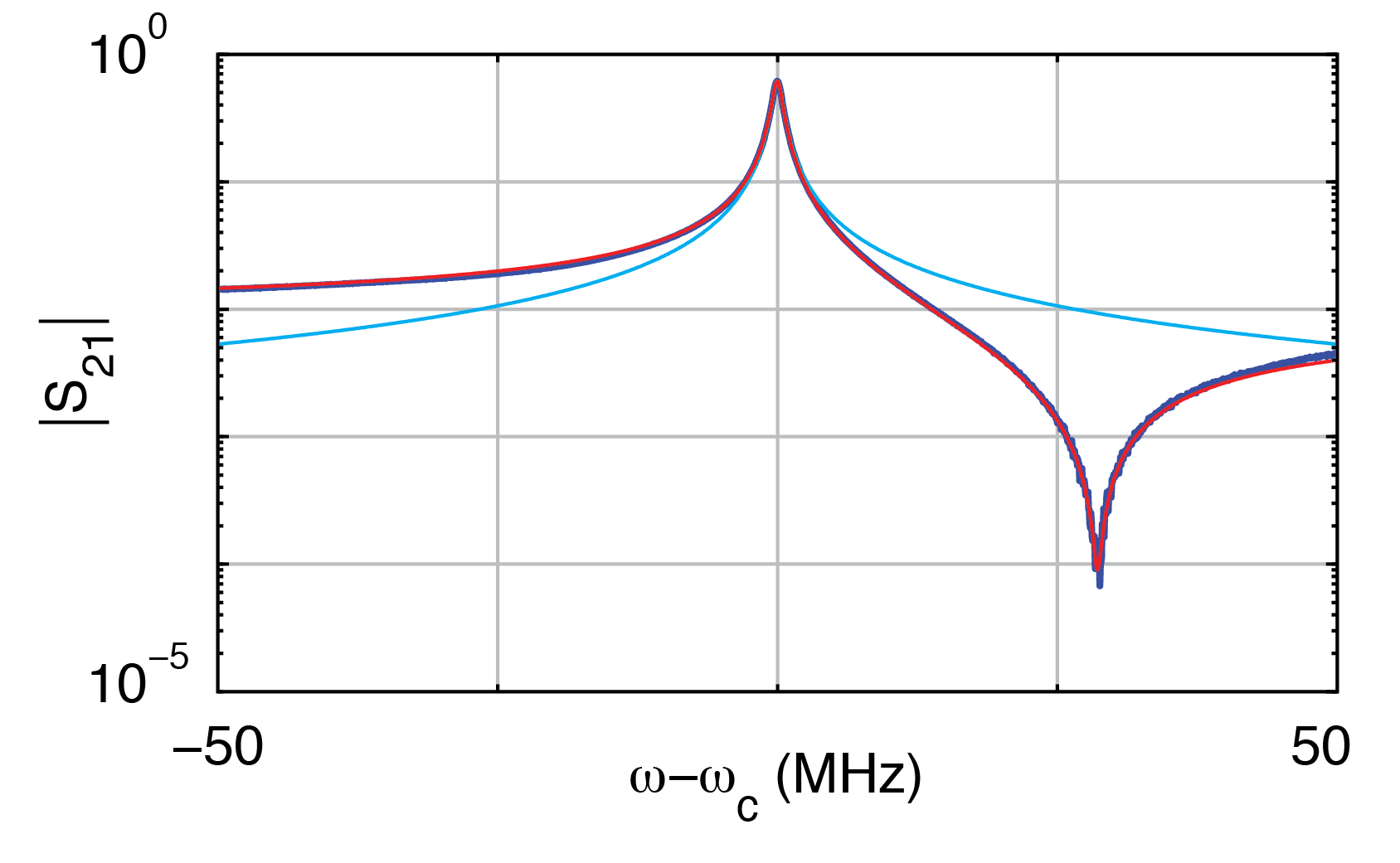}
\caption{Microwave driven response, $|S_{21}(\omega)|$ (dark blue), with shunt capacitor model fit,  $|S_{21}^{\rm s}(\omega)|$(red), and ideal model fit, $|S_{21}^0(\omega)|$ (light blue).}\label{Fig.FanoS21}
\end{figure}

We find that the driven response of the microwave circuit noticeably deviates from a Lorentzian lineshape at frequencies outside the resonator linewidth.  One distinct feature of the observed spectrum is an anti-resonance (Fig.\ref{Fig.FanoS21}),  indicating interference of multiple current channels at the output of the microwave circuit. As a first step to understand this behavior, we model the input and output transmission line discontinuities with shunt capacitors \cite{Pozar:2004}, $C_{s,\rm in}$ and $C_{s,\rm out}$, as presented in Fig. \ref{Fig.ShuntCircModel}.


We recover the circuit transmission, $S_{21}^s=\frac{V_{\rm amp}}{V_0}$, by applying Kirchoff's Circuit Law over the equivalent circuit model and solving the resulting system of equations.  Based on estimates of the circuit parameters, the transmission can be approximated as
\begin{align}
S_{21}^{\rm s}(\omega) = S_{21}^0(\omega)+2R_L\cdot j \omega_c C_{s,\rm out},
\end{align}
where $S_{21}^0(\omega)=\frac{-\sqrt{\kappa_R \kappa_L}}{j(\omega-\omega_c)+\kappa/2}$ is the Lorentzian transmission for the case of $C_{s,\rm in},C_{s,\rm out}\rightarrow 0$, $j=\sqrt{-1}$, and $R_L=50\Omega$ is the source impedance.  Notably, we also find that the voltage at the resonator, $V_{\rm smr}(\omega)$, is negligibly modified by these shunt capacitors.  This shows that the additional channels can be treated solely as modification to the cavity output scattering rate.

Fitting $S_{21}$ data from our device to this model, we estimate $C_{\rm out} = C_{s,\rm out} = 2.7$ fF.  We believe that these values are realistically acceptable given the geometry of our device and the proximity of the output coupler to both the tank circuit and the ground plane.  Additionally, shunt capacitance from wirebonds will also contribute to this effect.

From this fit, the ratio between transmission at negative and positive pump detunings is $2.4$ dB, showing close agreement with the ratio of $2.6$ dB observed in above calibrations.  From the form of $S_{21}^s$, we pull out the correction factor described in Sec.3.  Specifically, 
\begin{equation}
\cavSkew(\omega) = 4 \, R_L \, \omega_c C_{s,out} \, \frac{\kappa}{\sqrt{\kappa_L \kappa_R}} \left(\frac{\omega-\omega_c}{\kappa}\right).
\end{equation}
Since $\frac{\cavSkew(\omega_-)}{\cavSkew(\omega_+)}=-1$ and $\frac{1+\cavSkew(\omega_-)}{1+\cavSkew(\omega_+)}=2.6$ dB from measurement, we recover $\cavSkew(\omega_-) = 0.29$.

